\documentclass[prc,preprint,showpacs,floatfix,nofootinbib,tightenlines]{revtex4}
\usepackage{graphicx}  %
\usepackage{bm}  %

\usepackage{amssymb}

\begin{document}

\newcommand{\alphabar}{\overline\alpha}
\newcommand{\alphabarp}{\overline\alpha\,{}'}
\newcommand{\Azero}{A_0}

\newcommand{\bzero}{b_0}

\newcommand{\Cs}{C_{\rm s}}
\newcommand{\cs}{c_{\rm s}}
\newcommand{\Cv}{C_{\rm v}}
\newcommand{\cv}{c_{\rm v}}

\newcommand{\dalem}{\frame{\phantom{\rule{8pt}{8pt}}}}
\newcommand{\del}{\partial}
\newcommand{\delN}{{\wt\partial}}
\newcommand{\Deltaevac}{\Delta{\cal E}_{\rm vac}}

\newcommand{\ed}{{\cal E}}
\newcommand{\edk}{{\cal E}_k}
\newcommand{\edkzero}{{\cal E}_{k0}}
\newcommand{\edv}{{\cal E}_{\rm v}}
\newcommand{\edvphi}{{\cal E}_{{\rm v}\Phi}}
\newcommand{\edvphizero}{{\cal E}_{{\rm v}\Phi 0}}
\newcommand{\edzero}{{\cal E}_{0}}
\newcommand{\Efermistar}{E_{{\scriptscriptstyle \rm F}}^\ast}
\newcommand{\Efermistarzero}{E_{{\scriptscriptstyle \rm F}0}^\ast}
\newcommand{\etabar}{\overline\eta}
\newcommand{\ezero}{e_0}

\newcommand{\fomega}{f_\omegav}
\newcommand{\fpi}{f_\pi}
\newcommand{\fv}{f_{\rm v}}
\newcommand{\fvt}{\widetilde\fv}

\newcommand{\gA}{g_A}
\newcommand{\gammazero}{\gamma_0}
\newcommand{\gomega}{g_\omegav}
\newcommand{\gpi}{g_\pi}
\newcommand{\grho}{g_\rho}
\newcommand{\grad}{{\bm{\nabla}}}
\newcommand{\gs}{g_{\rm s}}
\newcommand{\gv}{g_{\rm v}}

\newcommand{\fm}{\mbox{\,fm}}

\newcommand{\infm}{\mbox{\,fm$^{-1}$}}
\newcommand{\isovectorTensor}{s_{\tauvec}}
\newcommand{\isovectorTensorN}{\wt\isovectorTensor}
\newcommand{\isovectorVector}{j_{\tauvec}}
\newcommand{\isovectorVectorN}{\wt j_{\tauvec}}

\newcommand{\kappabar}{\overline\kappa}
\newcommand{\kfermi}{k_{{\scriptscriptstyle \rm F}}}
\newcommand{\kfermizero}{k_{{\scriptscriptstyle \rm F}0}}
\newcommand{\Kzero}{K_0}

\newcommand{\LdotS}{\bm{\sigma\cdot L}}
\newcommand{\lsim}{\lower0.6ex\vbox{\hbox{$\ \buildrel{\textstyle <}
         \over{\sim}\ $}}}
\newcommand{\lzero}{l_{0}}

\newcommand{\Mbar}{\overline M}
\newcommand{\Mbarzero}{\Mbar_0}
\newcommand{\MeV}{\mbox{\,MeV}}
\newcommand{\momega}{m_\omegav}
\newcommand{\mpi}{m_\pi}
\newcommand{\mrho}{m_\rho}
\newcommand{\ms}{m_{\rm s}}
\newcommand{\Mstar}{M^\ast}
\newcommand{\Mstarzero}{M^\ast_0}
\newcommand{\mv}{m_{\rm v}}
\newcommand{\mzero}{{\rm v}_{0}}

\newcommand{\Nbar}{\overline N}

\newcommand{\omegaV}{V}
\newcommand{\omegav}{{\rm v}}

\newcommand{\Phizero}{\Phi_0}
\newcommand{\psibar}{\overline\psi}
\newcommand{\psidagger}{\psi^\dagger}
\newcommand{\pvec}{{\bf p}}

\newcommand{\rhoB}{\rho_{{\scriptscriptstyle \rm B}}}
\newcommand{\rhoBt}{\wt\rho_{{\scriptscriptstyle \rm B}}}
\newcommand{\rhoBzero}{\rho_{{\scriptscriptstyle \rm B}0}}
\newcommand{\rhominus}{\rho_{-}}
\newcommand{\rhoplus}{\rho_{+}}
\newcommand{\rhos}{\rho_{{\scriptstyle \rm s}}}
\newcommand{\rhospzero}{\rho'_{{\scriptstyle {\rm s} 0}}}
\newcommand{\rhost}{\wt\rho_{{\scriptstyle \rm s}}}
\newcommand{\rhoszero}{\rho_{{\scriptstyle {\rm s}0}}}
\newcommand{\rhotau}{\rho_{\tauvec}}
\newcommand{\rhotaut}{\wt\rho_{\tauvec}}
\newcommand{\rhothree}{\rho_{3}}
\newcommand{\rhothreet}{\wt\rho_{3}}
\newcommand{\rhozero}{\rho_0}

\newcommand{\scalar}{\rhos}
\newcommand{\scalarN}{{\rhost}}
\newcommand{\Szero}{S_0}

\newcommand{\tauvec}{{\bm{\tau}}}
\newcommand{\tauthree}{\tau_3}
\newcommand{\Tr}{{\rm Tr\,}}

\newcommand{\umu}{u^\mu}
\newcommand{\Ualpha}{U_{\alpha}}
\newcommand{\Ueff}{U_{\rm eff}}
\newcommand{\Uzero}{U_0}
\newcommand{\Uzerop}{U_0'}
\newcommand{\Uzeropp}{U_0''}

\newcommand{\vecalpha}{{\bm{\alpha}}}
\newcommand{\veccdot}{{\bm{\cdot}}}
\newcommand{\vecnabla}{{\bm{\nabla}}}
\newcommand{\vecpi}{{\bm{\pi}}}
\newcommand{\vectau}{{\bm{\tau}}}
\newcommand{\vectorj}{j_{\scriptscriptstyle V}}
\newcommand{\vectorN}{{\wt\vector}}
\newcommand{\vecx}{{\bf x}}
\newcommand{\Vopt}{V_{\rm opt}}
\newcommand{\Vs}{V_s}
\newcommand{\Vzero}{V_0}

\newcommand{\wt}{\widetilde}
\newcommand{\wzero}{w_0}
\newcommand{\Wzero}{W_0}

\newcommand{\xvec}{{\bf x}}

\newcommand{\zetabar}{\overline\zeta}
%
%
%
% End: Simple substitution macros
%

% other definitions
\newcommand{\beq}{\begin{equation}}
\newcommand{\eeq}{\end{equation}}
\newcommand{\beqa}{\begin{eqnarray}}
\newcommand{\eeqa}{\end{eqnarray}}

% uncomment \draft to have PACS numbers appear

\title{Neutron Radii in Mean-Field Models}

\author{R. J. Furnstahl}\email{furnstahl.1@osu.edu}
\affiliation{Department of Physics \\
         The Ohio State University,\ \ Columbus, OH\ \ 43210}

%
%\date{\today}
\date{December, 2001}

\begin{abstract}
Bulk nuclear observables
such as charge radii and binding energies are well described
by both nonrelativistic and covariant
mean-field models. 
However, predictions of neutron radii, which are not tightly constrained
by reliable data, vary significantly.
The nature of this variation is investigated using correlations
between basic properties of the models and the neutron skin thickness in 
lead.
The results suggest that conventional covariant models are too limited.
The study is guided by principles and insights of effective field theory (EFT), 
such as power counting,
and the relation of mean-field models to a more general EFT approach
to nuclei is discussed.
\end{abstract}

\smallskip

\maketitle

\section{Introduction}
\label{sec:introduction}

Bulk nuclear observables
such as charge radii and binding energies are well described
by both nonrelativistic and covariant
mean-field models \cite{BROWN98,RING96}.
Yet the ``best fit'' calculations of each type
in the literature disagree substantially on 
neutron radii \cite{HOROWITZ01,BROWN98,RING96}.
The spread in mean-field predictions and a recent analysis of
experimental determinations  \cite{HOROWITZ01}
suggest that neutron radii in medium to
heavy nuclei are inadequately determined at present by
either theory or experiment.
Better values for neutron radii could have widespread impact,
from reducing  uncertainties in atomic parity 
violation experiments \cite{HOROWITZ01}  to constraining
properties of the surface crust
of neutron stars \cite{HOROWITZ00}.
More generally,
the lack of control over neutron radius predictions implies
large uncertainties in extrapolations to neutron- or proton-rich nuclei,
which play an important role in nuclear astrophysics \cite{WP2000}.

There are arguments in the literature that mean-field models of nuclei are
inadequate because  the mean-field approximation is not valid at
ordinary nuclear densities or that essential correlations are
excluded \cite{AKMAL98}. 
However, these are misleading arguments, which implicitly assume that
one is performing
a Hartree or Hartree-Fock calculation of some underlying interaction.
In fact, it is more appropriate to identify these calculations with Kohn-Sham
density functional theory (DFT)  \cite{KOHN65,DREIZLER90,ARGAMAN00}, 
in which the full Hartree plus
exchange-correlation functionals (and not Hartree alone!) are approximated by a
parametrized form.%
\footnote{The observation that mean-field models can be interpreted
as Kohn-Sham functionals
has been made by Brack for Skyrme models \cite{BRACK85} and by
Schmid et al.\ for covariant models \cite{SCHMID95,SCHMID95a}.}
The important observation is that mean-field
models are specified by a {\em universal\/} energy functional
of the density,
which is minimized iteratively for each nucleus through the introduction
of auxiliary single-particle orbitals.
This is the structure of a Kohn-Sham DFT.
The solution procedure takes a small fraction
of the computational effort of conventional direct many-body solutions. 
The universality is important because the difficulties in finding
solutions to many different nuclei are shifted to
constructing the functional once; if the functional is easily
evaluated then the cost of subsequent applications to
each additional finite nucleus is small, and one can 
scale to large numbers of nucleons.
  
``Mean field'' in this context really implies a 
limited form of the analytic and nonlocal structure that can appear
in the exact energy functional.
We interpret 
the mean-field models as a form of Kohn-Sham DFT in which the functional is
approximated by a type of generalized gradient 
expansion \cite{PERDEW96,PERDEW99}.
From this point of view, the underlying framework is completely general and the
limitations of present-day mean-field calculations 
could be systematically removed
without losing the computational advantages of the approach.
{\em Deficiencies in the mean-field model predictions should therefore be traced
to deficiencies in the approximate functional rather than to a breakdown
of the underlying framework.\/}

In this paper,
we adapt this philosophy to a study of mean-field neutron radii. 
The nature of the variation in radii is investigated using correlations
between basic characteristics of the models and the predicted
neutron skin thickness in lead.
In this way, we can identify how properties of the mean-field
energy functionals are related to observables and identify possible
deficiencies in the functional.
The study is guided by the principles and insights of effective field theory 
(EFT),  with an eye on the ultimate goal of embedding
mean-field models into a more general DFT/EFT approach to 
nuclei \cite{FURNSTAHL00}.
With an EFT approach one can exploit
{\em power counting\/} 
to systematically approximate the energy functional.
In the present context, 
power counting means that one can estimate the contribution 
to the energy of
individual terms in the energy functional, even those excluded by the
model truncation. 
The promise of this approach is supported by 
empirical evidence that EFT power counting in mean-field models is 
robust \cite{FURNSTAHL99b,FURNSTAHL97,RUSNAK97,FURNSTAHL97c}.  

In principle,
a detailed analysis {\em during the fitting process\/} would best reveal
how observables such as the neutron radius
are correlated with other properties of the model.
As an alternative, we consider a large set of models all fit to roughly
the same data at about the same accuracy, 
and then examine the correlations graphically.
This strategy has been used with covariant models
to study deformations in nuclei \cite{FURNSTAHL87} and
spin-orbit splittings \cite{FURNSTAHL98}.
Brown has used such correlations with the Skyrme model recently to 
address many of the same issues about neutron radii \cite{BROWN00}
and Typel and Brown have extended these results to conventional
covariant models \cite{BROWN01}.
In our analysis,
we use conventional parameter sets from the literature but also
generalized functionals developed for EFT studies.
The latter
provide a wide class
of models that provide very good fits to the usual bulk nuclear
data but which also span
a wide range of observables that are not well constrained by this data. 

We find that the neutron skin thickness is dominantly correlated with 
the bulk symmetry energy and its density dependence.  
Thus the properties of asymmetric infinite nuclear matter (as opposed to
surface properties) are most important.
Our conclusions on the physics determining the skin are in accord
with many of the observations in
Ref.~\cite{OYAMATSU}, which are based on applying a macroscopic model
and comparing to two particular mean-field models, and by Brown \cite{BROWN00}
and Brown and Typel \cite{BROWN01}.
(Note: many of the same issues were discussed 
long ago by Bodmer who used a semi-empirical statistical method
\cite{BODMER60}.  Other recent studies of the neutron skin in
mean-field models
include
Refs.~\cite{CHEN99,PATYK99,DOBACZEWSKI99,VRETENAR00,MIZUTORI00}.)
The differences between predictions from standard Skyrme and covariant
models can then be traced to differences in the density dependence
of the symmetry energy.  
A comparison of the symmetry energy derived from more microscopic
approaches that are tied to free-space nucleon-nucleon scattering 
%(matching the energy functional, in effect) 
implies that conventional
covariant models based on heavy meson interactions are too simple.

The plan of the paper is as follows.
In Sect.~\ref{sec:mean-field}, we describe the mean-field
functionals and and in Sect.~\ref{sec:power} the EFT power counting.
The correlation analysis is carried out in Sect.~\ref{sec:correlations} 
and the consequences discussed in Sect.~\ref{sec:discussion}.
Sect.~\ref{sec:summary} is a summary.

\section{Overview of Mean-Field Energy Functionals}\label{sec:mean-field}

There are two major classes of mean-field models in general use, 
which can be formulated in terms of nonrelativistic
(Skyrme) or covariant energy functionals.  The covariant models
can be further subdivided into meson and point-coupling models.
We will compare all three types in our correlation analysis.
These models and their implementations
are well documented in the literature 
\cite{KREWALD,FRIEDRICH86,CHABANAT97,POMORSKI97,BROWN98,SEROT86,RUFA88,%
SEROT92,RING96,SEROT97,FURNSTAHL97,LALAZISSIS97,NIKOLAUS97,RUSNAK97}
and we will only discuss
here details relevant to the present investigation.
We focus directly on energy 
functionals instead of Hamiltonians or Lagrangians 
to stress  the connection to a more general formulation
in terms of density functional theory.
The  usual construction of the functional starts by postulating 
interactions (either a Skyrme effective interaction or a Lagrangian) 
and then treats them in the Hartree approximation.%  
\footnote{Note that the Skyrme models actually apply a
Hartree-Fock approximation, but since the forces are zero range, this 
doesn't affect the form of terms in the energy functional.}

The Skyrme models are based 
on nonrelativistic zero-range, density-dependent 
interactions.  
The energy functional most commonly used in the present-day
literature has evolved only slightly 
from the original form proposed; an explicit expression for
the interaction is given
in Ref.~\cite{BROWN98}.
The corresponding functional can be written  
as powers and gradients
of the isoscalar density $\rho=\rho_p+\rho_n$ and isovector density
$\rho_3 = \rho_p-\rho_n$, the corresponding kinetic energy
densities $\tau$ and $\tau_3$,  and spin-orbit currents.
As such it can be considered a Kohn-Sham functional constructed
as a generalized gradient expansion \cite{PERDEW96,PERDEW99}.
Of particular note are terms proportional to
$\rho^{2+\alpha}$ and $\rho^{\alpha}\rho_3^2$, which in modern Skyrme parameter
sets appear with fractional $\alpha$.
Although extended Skyrme models have been considered \cite{KREWALD},
the density expansion stops with these terms.
There are usually four isovector parameters (which are linear combinations of
the traditional parameters $t_0$--$t_3$, $x_0$--$x_4$, and $W_0$
\cite{BROWN98}), corresponding to $\rho_3^2$,
$\rho^{\alpha}\rho_3^2$, $\rho_3 \tau_3$, and $(\nabla\rho_3)^2$ terms
(there are also isovector terms involving the spin-orbit
currents).
While the Skyrme interaction has been motivated in terms of underlying
nucleon-nucleon interactions \cite{NEGELE70,NEGELE72}, 
the connection has only been made
qualitatively and the parameters are treated phenomenologically.
A goal of the EFT/DFT program advocated here will be to revisit
this connection and seek systematic improvement.

Many Skyrme parameter sets are available in the 
literature \cite{FRIEDRICH86,CHABANAT97,POMORSKI97,BROWN00},
representing a wide range of fits to bulk nuclear properties,
although the fits are not all to the same set of observables and are not
of equal quality.
The Skyrme parameter set developed by Brown \cite{BROWN98} is fit to the
widest range of properties and 
serves as the quality-of-fit benchmark for this class of models.
We will use most of the sets from the literature plus many
new ones generated for this investigation.
The new sets are fit to binding energies, charge densities, and 
spin-orbit splittings of doubly magic nuclei
as described for covariant models in
Refs.~\cite{FURNSTAHL97} and \cite{RUSNAK97}.
However, a value for the neutron
radius in lead was also specified so as to generate sets that cover 
as large a range
in neutron radii as is compatible with a good fit to the other properties. 
Only new sets with a quality of fit
comparable to Ref.~\cite{BROWN98} were used in our analysis.

The most commonly used
covariant mean-field models are derived from Lagrangians
with nucleons coupled to heavy mesons
and are characterized by large, isoscalar, 
mean meson fields that are
several hundred MeV in magnitude \cite{SEROT86,SEROT92,RING96,SEROT97}.
Meson exchange in these models is spacelike and far off-shell, so
the corresponding parameters are at best only roughly related to the physical
meson spectrum.
We describe these energy functionals as ``covariant'' rather than
``relativistic,'' as is conventional in the literature, since the 
latter suggests
the importance of {\em kinematic\/} relativity---large velocities
and large $p/M$---whereas these corrections are small for nuclei.
Rather it is the {\em potentials\/} where relativity is most important, 
and this
is manifested in a Lorentz covariant representation, which maintains the 
distinction between Lorentz scalars and zeroth components of 
four-vectors \cite{FURNSTAHL00b}.  
(It is not Lorentz {\em invariant\/}, since the four velocity of a
nucleus or nuclear matter defines a preferred frame.)

In practice, the form of the energy functional has been
limited to terms that can arise
from the one-loop or Hartree approximation to a Lorentz invariant Lagrangian.
In this sense the functionals are less general than the Skyrme
functionals.
This comparison is somewhat misleading, however, since a functional
built from integral powers of
both Lorentz scalar and vector densities will, when reduced to
nonrelativistic form, contain terms non-analytic in the density
(i.e., analogous to the $\rho^{2+\alpha}$ term for particular values of
$\alpha$) \cite{FOREST95,SEROT97}.
There are several standard covariant-model
parameter sets in the literature 
\cite{RUFA88,LALAZISSIS97} and many more 
have been generated recently in the course of
EFT investigations \cite{FURNSTAHL97,RUSNAKTHESIS}.
{\em We use them all in the present investigation.\/}
In conventional mean-field parametrizations, 
the model is usually truncated to minimize the 
parameters needed
to provide a good fit, with the hope of maximizing predictive power.
Terms through the fourth power of the scalar mean field are included but only
through the second power of the other mean
fields; there is only one non-gradient
isovector parameter.  
In an EFT-inspired analysis, one uses power counting arguments
to organize the energy functional and reveal how many parameters can
be fixed from the data. 
The EFT parameter sets include more interaction terms but are still not
complete.  For example, a meson corresponding to isovector scalar
exchange was not considered because its effects were found to be small
in previous investigations.  In Ref.~\cite{LIU01}, this is argued to
have an important effect on the symmetry energy.

An alternative to explicit mesons are ``point coupling'' (PC) models,
which feature contact interactions and derivatives among nucleons
only rather than
meson exchange \cite{NIKOLAUS97,RUSNAK97,BURVENICH01}. 
The energy functional here is composed of powers and gradients of Lorenz scalar
($\rho_s$), vector ($\rho_B$) and tensor densities and currents
\cite{FURNSTAHL99b}.
Point coupling models are somewhat more flexible than covariant
meson models because the assumption of an underlying meson exchange
does not constrain terms in the energy functional.
Thus a term such as $(\nabla\rho_s)^2$ would have a
definite sign if based on an underlying meson exchange but could have
either sign in a point-coupling
model.

In both types of covariant models, 
the isovector interactions in conventional
formulations are quite limited, with the form of simple rho meson
exchange or its point-coupling equivalent.
Since the corresponding mass is largely unconstrained by fits to
nuclear properties, there is
really only one free parameter.  This is in contrast to conventional
Skyrme
interactions, which have up to four
parameters
controlling isovector contributions
(although two are largely irrelevant for the properties discussed
here \cite{BROWN98}).

\section{Power Counting in the Energy Functional}
\label{sec:power}

An effective field theory (EFT) describes low-energy physics with
low-energy degrees of freedom.
 Underlying EFT is the principle that while the short-distance,
ultraviolet behavior of the effective theory may be incorrect, 
it can be corrected systematically by the renormalization of local operators
(``counterterms'') \cite{LEPAGE89}.
There is considerable freedom in the choice of representation of these
operators; while one choice is not more ``correct'' than another,
it may be more convergent or easier to calculate.
It is in this context  that we compare nonrelativistic and covariant
energy functionals of the density for nuclei.
A summary of the EFT philosophy as applied to mean-field models is given in
Ref.~\cite{FURNSTAHL00}.  For our purposes, the most important concept
is that of power counting in the energy functional, which associates a
natural size to every term based on conjectured underlying scales.

Will an EFT approach be useful for heavy nuclei?
An EFT relies on the separation of scales, so we should identify
the momentum
scale $\Lambda$ that divides short from long-distance physics.
Long-distance physics must be treated explicitly while short-distance
physics can be reproduced with local operators.
For the latter
we might expect a density expansion in powers of $\kfermi/\Lambda$,
where $\kfermi$ is the Fermi momentum.
(This is manifest in the treatment of a dilute Fermi system
\cite{HAMMER00}, but
has not been shown for nuclear systems.)
There are two immediate problems.
First, nuclear matter equilibrium $\kfermi$ is significantly larger
than the pion mass, which implies that an explicit treatment
of pion physics as long-range physics is essential for a useful
EFT. 
Second, nuclear matter saturation might only occur when
$\kfermi \approx \Lambda$, because only then could different terms
in the density expansion compensate each other.  

However, the bulk physics of heavy nuclei is dominated by isoscalar
physics (see below), 
and the dominant pion contributions appear to be two-pion
physics, which is associated with the Lorentz scalar ``$\sigma$'' meson
\cite{SEROT92}. 
The success in using $\sigma$ exchange in two-nucleon potentials
implies that 
the low-energy tail of the two-pion contribution
is not well resolved, particularly in nuclei.%
\footnote{Indeed, one must work very carefully to discern the
long-range chiral two-pion-exchange tail in the detailed two-nucleon
scattering data \cite{RENTMEESTER99}.}
The consequence is that in an EFT,
nuclear structure is dominated by short-distance
contact terms (``low-energy constants''), which implies that a model with
$\Lambda \approx 600\,$MeV without explicit pions is actually a good
starting point.
This argument also implies that isovector observables such as the
neutron skin, which could be sensitive to one-pion-exchange physics, 
may be inadequately described.

The potential problem with applying EFT to saturating systems,
in which saturation is only possible at the breakdown scale, is apparently not
realized in nuclei.  In fact, by conventional analysis,
equilibrium nuclear matter is an
anomalously dilute Fermi system.
One such analysis, by
Jackson \cite{JACKSON92,JACKSON94},
is based on saturation driven by a strongly repulsive short-range
interaction (which we'll call a ``hard core'') with range $c$
(so that $\Lambda \approx 1/c$).
The size $R$ of a saturating system scales with the number of particles
$A$ to the one-third power, which defines a characteristic length scale
$r_0$:
\beq
    R \sim r_0 A^{1/3}  \ .
\eeq
Jackson estimates the limits of saturation to be bounded by
$0.552c \leq r_0 \leq 2.4c$ \cite{JACKSON92}.  
This estimate is consistent with 
liquid ${}^3$He, where $r_0 \approx 0.96c$, but with a nuclear
hard core at $c=0.4\,$fm, $r_0$ is actually beyond the bound:
$r_0 \approx 2.75c$, which implies that saturation in nuclei is
unconventional.
One possible explanation for saturation in an EFT with a density
expansion is that two low orders might largely cancel because of
an anomalously small coefficient multiplying a leading term, while all higher
orders follow the expansion in $\kfermi/\Lambda$.  This is the
mechanism for saturation in covariant mean-field models \cite{FURNSTAHL00}.

We assume that a counting consistent with low-energy QCD, including
pions as light degrees of freedom, is appropriate for our 
mean-field energy functionals.
Georgi-Manohar
naive dimensional analysis, or NDA, assigns appropriate powers of the
pion decay constant $f_\pi\approx 93\,$MeV 
and a scale of non-goldstone boson physics
$\Lambda$
(about 600\,MeV) to terms in a low-energy effective Lagrangian
of QCD \cite{GEORGI84b,GEORGI93b}.
The counting for 
a generic term in a covariant meson Lagrangian is found from
\beq
    \beta\,  [\Lambda^4/g^2]
    \biggl( {g^2\overline \psi\psi\over \Lambda^3} \biggr)^l
    \biggl( {\nabla\over \Lambda} \biggr)^p
    {1 \over m!}\biggl(  {g\phi\over \Lambda} \biggr)^m
    {1\over n!}\biggl(  {gV\over \Lambda} \biggr)^n
\eeq
where $g \sim \Lambda/f_\pi$    
and Dirac and isospin operators are not shown.
(See Ref.~\cite{FURNSTAHL97} for a discussion of power counting with
covariant meson Lagrangians including pions.)
The assumption here is that this power counting, intended for terms in a
Lagrangian, can be applied directly to the energy functional through
the associations $\rhos \leftrightarrow \overline\psi\psi$,
$\rhoB \leftrightarrow \psidagger\psi$ and so on.  
This connection is immediate if the
mean-field model
is viewed as being only a one-loop calculation but has not been fully
clarified in the more general context of DFT.
Skyrme functionals follow an analogous power counting behavior
with powers of $\rho$ and $\tau$ counted using
$\rho\leftrightarrow \psidagger\psi$, $\tau\leftrightarrow
(\nabla\psi)^2$, and so on (see Ref.~\cite{FURNSTAHL97c} for
details).

%%%%%%%%%%%%%%%%%%%%%%%%%%%%%%%%%%%%%%%%%%%%%%%%%%%%%%%%%%
\begin{figure}[t]
\begin{center}
\includegraphics[width=4.1in,angle=0,clip=true]{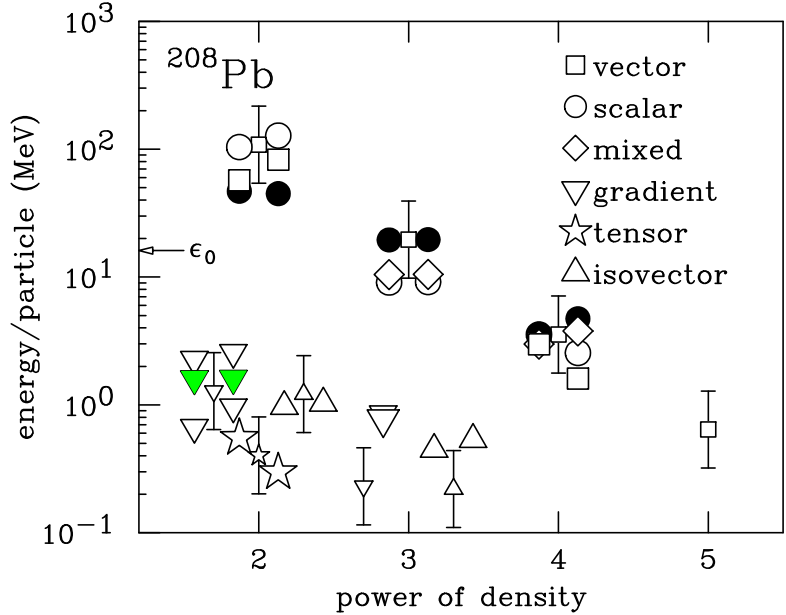}
\caption{Contributions to the energy per particle in ${}^{208}$Pb
 for two covariant point-coupling models from Ref.~\protect\cite{RUSNAK97}.
 Absolute values are shown and
 the filled symbols are net values for the sum of scalar and vector
 terms. 
The small symbols indicate estimates based on NDA
[Eq.~(\protect\ref{eq:estimate})], with the error bars
corresponding to natural coefficients from 1/2 to 2.}
\label{fig:pb208energy1}
\end{center}
\end{figure}
%%%%%%%%%%%%%%%%%%%%%%%%%%%%%%%%%%%%%%%%%%%%%%%%%%%%%%%%%%%%%%

If we have identified the appropriate underlying length scales,
naturalness
implies that the dimensionless constant $\beta$ is of order unity.
Given this result, we can estimate contributions to the energy
from any term in the mean-field energy functional.
For example, a local-density 
estimate (per nucleon) for a general isoscalar term 
with dimensionless coefficient $\beta$ can be made according to
\beq
   {1\over A}\fpi^2\Lambda^2 \int\! d^3x\, \beta\, (\rhost)^i (\rhoBt)^j
      \approx \beta \Lambda 
	 \left( { \langle \rhoB \rangle \over \fpi^2\Lambda } \right)^{i+j-1}
     \ ,  \label{eq:estimate}
\eeq
where $\rhost$ and $\rhoBt$ are naturally scaled scalar and baryon
densities (e.g., $\rhost\equiv \rho_s/\fpi^2\Lambda$)
and $\langle \rhoB \rangle$ is the average density in the nucleus
being considered.
We take $1/2 < \beta < 2$ as a reasonable range of natural coefficients. 
Other local density prescriptions are detailed in
Refs.~\cite{FURNSTAHL99b} and \cite{FURNSTAHL98}.
Figure~\ref{fig:pb208energy1} 
shows these natural estimates with the associated error bars from the 
variation
in $\beta$ as small squares and triangles for ${}^{208}$Pb.
Actual contributions to the energy per particle in lead for two
covariant
point-coupling models that fit nuclear properties
are obtained by applying the Hellmann-Feynman theorem
\cite{FURNSTAHL99b}.  
The NDA estimates are
validated by these results and many other studies
\cite{FRIAR96b,FRIAR96,NIKOLAUS97,RUSNAK97,FURNSTAHL97c,FURNSTAHL97,BURVENICH01}. 

We will take as a guiding principle that NDA and naturalness are valid
for mean-field energy functionals.
Indeed, {\em all\/} of the best-fit models exhibit naturalness in all
or almost all coefficients.
A comparison of goodness-of-fit with this counting at different levels
of truncation reveals the limit
of the energy in constraining terms.
In particular, contributions of order 1\,MeV are barely constrained, so
contributions from 
beyond the fourth power of isoscalar densities or meson mean fields
are not distinguished from numerical noise \cite{FURNSTAHL99b}.
These results explain why the conventional covariant mean-field truncation
is phenomenologically successful.
In Ref.~\cite{FURNSTAHL99b}, a complete analysis like this
concluded that only six constants in covariant energy functionals are well
determined by fits to the usual bulk properties of nuclei.
These includes four isoscalar terms, one gradient term, and one
isovector term.
Brown  concludes that a similar number of parameters
are determined in fits to Skyrme models that use only the standard
observables \cite{BROWN98}.

The conventional set of observables used to fit parameters directly
constrains the isovector contribution only through the binding energy
(i.e., the symmetry energy).
Isovector contributions can be reliably estimated using
$\langle\rho_3\rangle \approx (Z-N)/A \times \langle\rhoB\rangle$.
The leading isovector contribution in Fig.~\ref{fig:pb208energy1}
is therefore suppressed relative to
the leading isoscalar contribution by the factor
${[(Z-N)/2A]^2}$ (the extra 1/2 comes from $\frac12\tau_3$
\cite{FURNSTAHL99b}), 
which is small even in lead.
This puts the leading term on the order of 1--2\,MeV,
which is just resolved by the fits ($\rhoB^5$ isoscalar
contributions are not resolved).
When more than one parameter is available, as shown in the
figure, only a linear combination is determined, and the subleading
coefficient is borderline unnatural \cite{RUSNAK97}.
While individual values for these couplings vary widely between
different parameter sets, the linear combination is well determined. 
The power counting analysis supports 
previous observations in meson models that,
at the mean-field level, the role of the $\rho$ meson is simply
to adjust the symmetry energy  \cite{HOROWITZ81,MUELLER96}.

We emphasize that despite the phenomenological success 
of the power counting, we do not as yet have a systematic EFT expansion
for the energy functionals.
In particular,
long-range loop contributions may lead to essential nonlocalities
and nonanalyticities in the energy functional. 
An underlying assumption that we are testing
is that long-distance contributions do not
disturb the power counting.
One should not assume, however, that long distance degrees of freedom
imply that the functional cannot take the form of powers of densities
and gradients.
For example, we note that Coulomb systems, which depend entirely 
on long range physics, 
are well
described by energy functionals based on
local density and gradient expansions \cite{ARGAMAN00}.

\section{Correlation Analysis}\label{sec:correlations}

Most observables in finite nuclei have a highly nonlinear and
correlated dependence on the parameters of the energy functional.
Predicting the correlations between features of the mean-field
functional and observables such as the neutron radius or the neutron
skin thickness is prone to error and plausible explanations often turn out
to be incorrect (as illustrated below for finite range effects).  
This is where power counting and EFT
arguments can be illuminating, even in the absence of a systematic
EFT expansion.
One can also go wrong by considering only a few ``best fit'' parameter
sets, which can lead to false conclusions because of the limited sampling.

Our strategy is to use many parameter sets, each a good fit to 
bulk properties of doubly magic nuclei.
We consider generalized versions of each type of mean-field model 
and look at
correlations between observables and characteristics of the models.
In order to fill in gaps that can obscure trends,
we also force particular neutron radii, being careful to keep only
sets with acceptably good fits to the standard properties.
We leave open the question of whether other finite nucleus observables
(such as inelastic scattering to collective states)
can further restrict the acceptable sets.

In the correlation plots that follow, the shape of a symbol indicates the type
of mean-field model:  circles for Skyrme, squares for covariant meson,
and triangles for covariant point coupling.
The shading indicates the origin of the model.
Black filled symbols are standard parameter sets from the
literature
\cite{FRIEDRICH86,CHABANAT97,POMORSKI97,BROWN00,RUFA88,LALAZISSIS97,NIKOLAUS97}.
Grey filled symbols are parameter sets generated for EFT
investigations \cite{FURNSTAHL97,RUSNAK97} and diagonal striped symbols are
unpublished generalized EFT sets from Ref.~\cite{RUSNAKTHESIS}.
Black and white checkered symbols are new parameter sets generated
for this investigation.
Other than this (partial) identification we treat all sets democratically.
{\em Note that not all models are plotted on all figures.\/}

%%%%%%%%%%%%%%%%%%%%%%%%%%%%%%%%%%%%%%%%%%%%%%%%%%%%%%%%%%
\begin{figure}[p]
\includegraphics[width=3.52in,angle=0,clip=true]{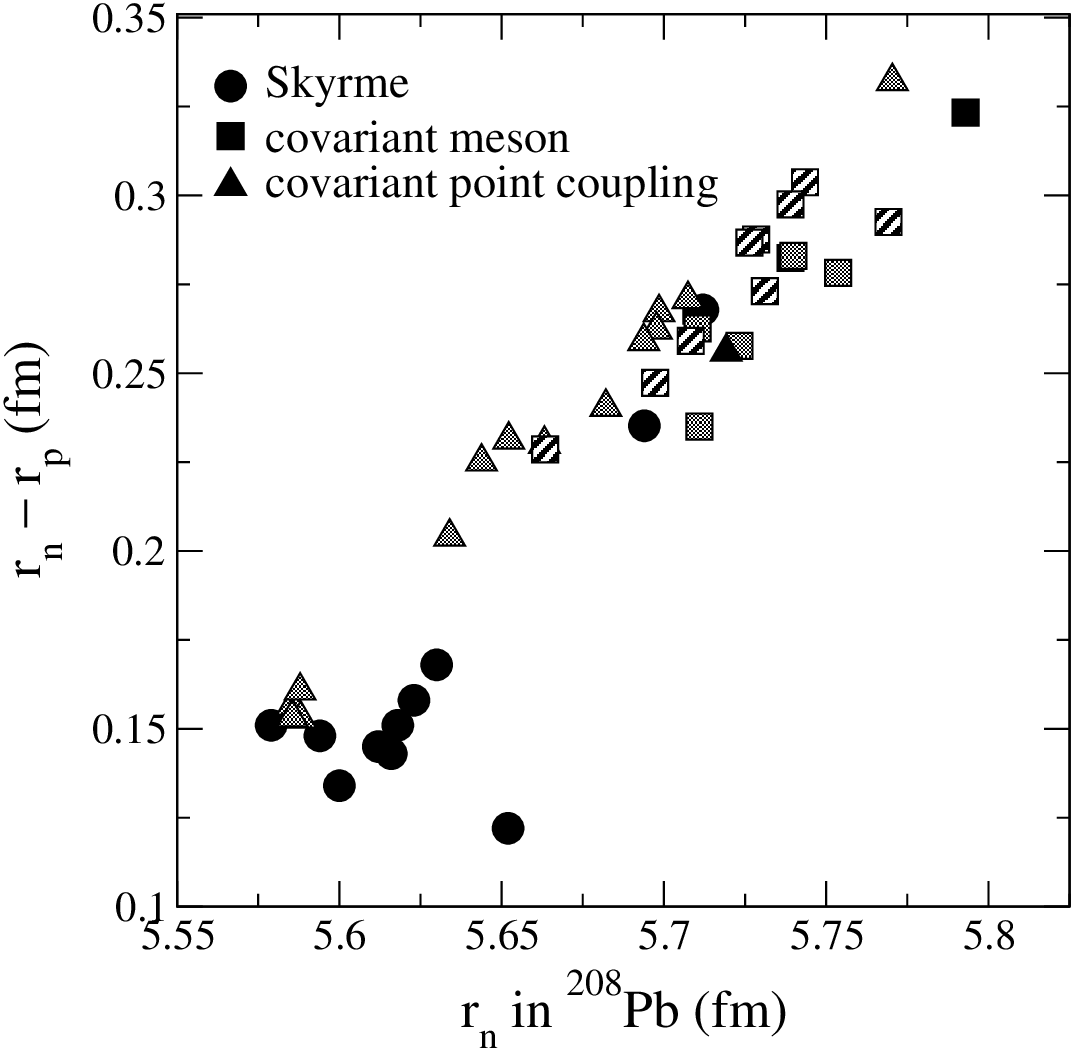}
\caption{Neutron skin thickness vs.\ neutron radius for 
a wide variety of mean-field models, as described in the text.
The shading indicates the origin of the model.
Black filled symbols are standard parameter sets from the
literature
\cite{FRIEDRICH86,CHABANAT97,POMORSKI97,BROWN00,RUFA88,LALAZISSIS97,NIKOLAUS97}.
Grey filled symbols are parameter sets generated for EFT
investigations \cite{FURNSTAHL97,RUSNAK97} and diagonal striped symbols are
unpublished generalized EFT sets from Ref.~\cite{RUSNAKTHESIS}.
%Black and white checkered symbols are new parameter sets generated
%for this investigation.
}
\label{fig:rnrprn}
%
%\vfill
\vspace*{.1in}
\includegraphics[width=3.52in,angle=0,clip=true]{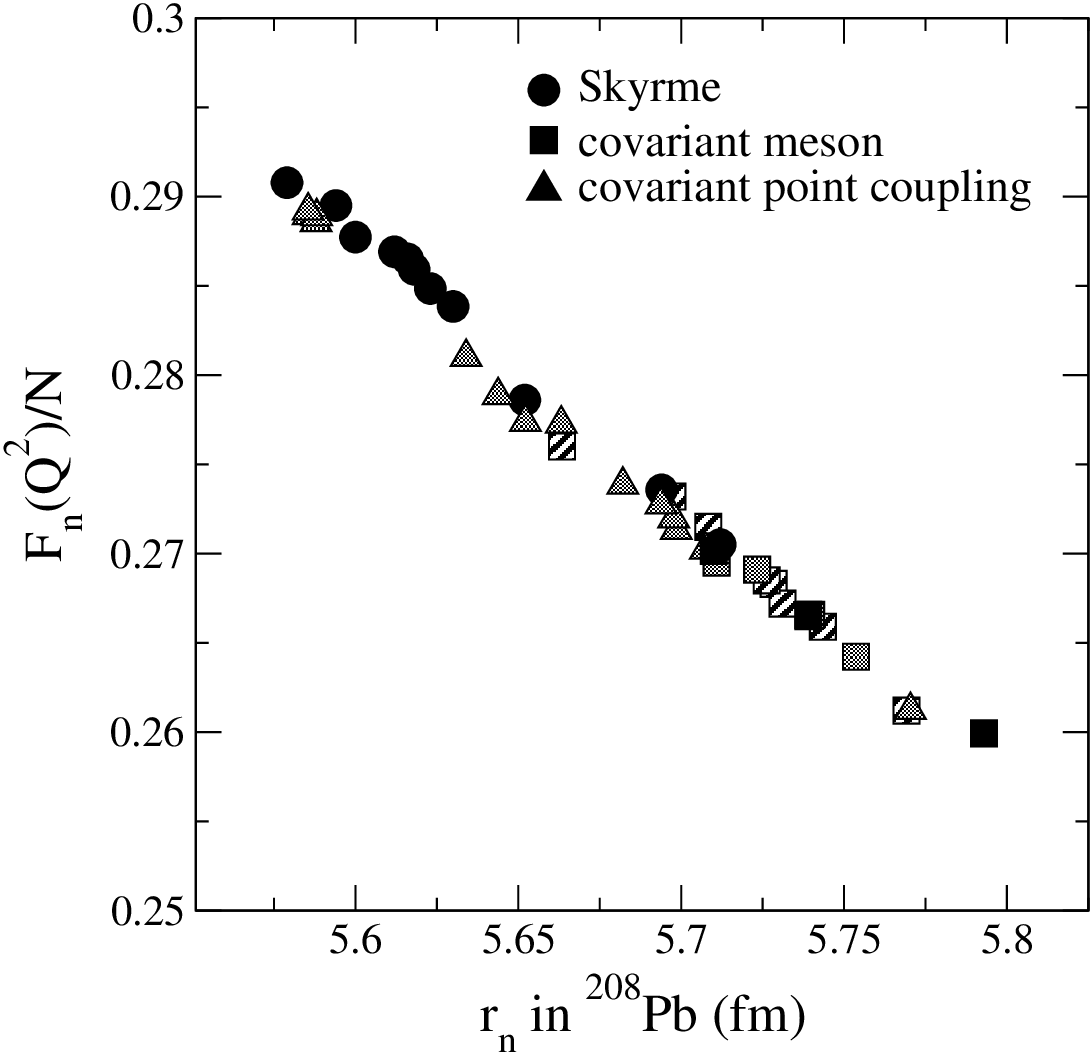}
\caption{The calculated neutron form factor (as defined in
Ref.~\cite{HOROWITZ01}) for $^{208}$Pb at
momentum transfer $Q = 0.45\,\mbox{fm}^{-1}$ vs.\
the point neutron radius in $^{208}$Pb for
a wide variety of Skyrme and covariant
mean-field models, as described in the text and in the caption
to Fig.~\ref{fig:rnrprn}.}
\label{fig:ffvsrn}
\end{figure}
%%%%%%%%%%%%%%%%%%%%%%%%%%%%%%%%%%%%%%%%%%%%%%%%%%%%%%%%%%%%%%

Comparisons between neutron radii or neutron skins from any given Skyrme model
and any given covariant  model
over a wide range of nucleon number $A$ show that the differences are very
systematic.
Fluctuations in the difference for particular nuclei, which
are affected by details such as the treatment of pairing in open-shell
nuclei, are small compared to the difference itself 
\cite{POMORSKI97,HOROWITZ01}. 
We conclude that understanding the skin thickness for one convenient
nucleus is sufficient at the resolution of this investigation.
We focus on $^{208}$Pb since it
has the largest skin of the doubly magic nuclei.

The skin thickness is a more robust isovector observable than
the neutron radius alone.
In principle, the charge radius of Pb$^{208}$ is very well determined
as are the factors that ``remove'' the charge form factor of the
proton (and other smaller corrections).  Thus, the point proton
radius $r_p$ should be very well fixed, and therefore 
the skin thickness should be closely correlated to $r_n$.  
In practice,
when $r_n-r_p$ is plotted as a function of $r_n$ for conventional
models,
there is a considerable spread, as seen in Fig.~\ref{fig:rnrprn}.
This spread reflects slightly different  form factor corrections
used in the models
and differences in how well $\langle r^2 \rangle_{\rm ch}$ 
in $^{208}$Pb is
reproduced by the fits.
We want to isolate isovector properties, 
so the skin thickness is more informative than the
individual radii.
Since the neutron radii are not included in the fits determining
these sets, the scatter is primarily driven by the binding energies (which
reflect the bulk symmetry energy).

Thus, we focus on the neutron skin thickness, $(r_n-r_p)$, in $^{208}$Pb
as a representative isovector ``observable'' for our study.
We put observable in quotes here because, in fact, 
neither the point proton radius, the point
neutron radius, nor their difference is actually an experimental observable.
Although it is standard practice,
the ``unfolding'' of a point proton radius from a charge radius
measurement is necessarily model dependent.
[See Ref.~\cite{FURNSTAHL01a} for a discussion from the EFT point of view
of the ambiguities in other quantities (occupation numbers) 
that are often treated as observables.]
We will assume that the natural size of this ambiguity  
is small compared to the size of the difference in neutron radii,
but this should be investigated further.
A consistent EFT
framework will obviate the need to discuss point data,
as illustrated in Ref.~\cite{FURNSTAHL97}, where nucleon charge form
factors are built into the Lagrangian and energy functional.

We also note that
the neutron radius is very highly correlated with the neutron form factor
at low momentum transfer.
This correlation is important for a proposed parity violation
experiment at Jefferson Lab, which seeks to make a five percent
measurement of the form factor at a momentum transfer of 0.45\,fm$^{-1}$.
An analysis of the experiment is given by Horowitz et al.\
\cite{HOROWITZ01}.
The relationship between the form factor at this momentum transfer
and the neutron radius in $^{208}$Pb is shown in Fig.~\ref{fig:ffvsrn}; 
it is evident that
the proposed measurement will directly constrain the neutron radius.

These figures already show that conventional covariant meson models 
(black squares) in
general predict larger skin thicknesses than conventional Skyrme
models  (black circles). 
A plausible explanation for the larger neutron skin thickness in
meson models compared to Skyrme is the finite range of the
interaction in the former.
The argument is that finite range meson exchange enables the
neutron density to extend further than the proton density
and still feel the
attraction from the protons.
However, this picture is not consistent with an EFT viewpoint.
The resolution scale associated with meson exchange in the covariant
models is $\Lambda \approx 600\,$MeV.
For momenta small compared to this scale, the details of the meson
exchange and the substructure of the hadrons are not resolved.
As a result, this short-range physics 
can be incorporated into the coefficients of
operators organized as a derivative expansion.
The consequence for low-momentum properties such as the rms radii
is that one cannot distinguish interactions with explicit ranges
of order $\Lambda^{-1}$ from interactions composed of
$\delta$-functions plus derivatives.
In principle, one needs an infinite number of terms for complete equivalence;
in practice power counting tells us we can truncate sharply.

%%%%%%%%%%%%%%%%%%%%%%%%%%%%%%%%%%%%%%%%%%%%%%%%%%%%%%%%%%
\begin{figure}[p]
\begin{center}
\includegraphics[width=4.0in,angle=0,clip=true]{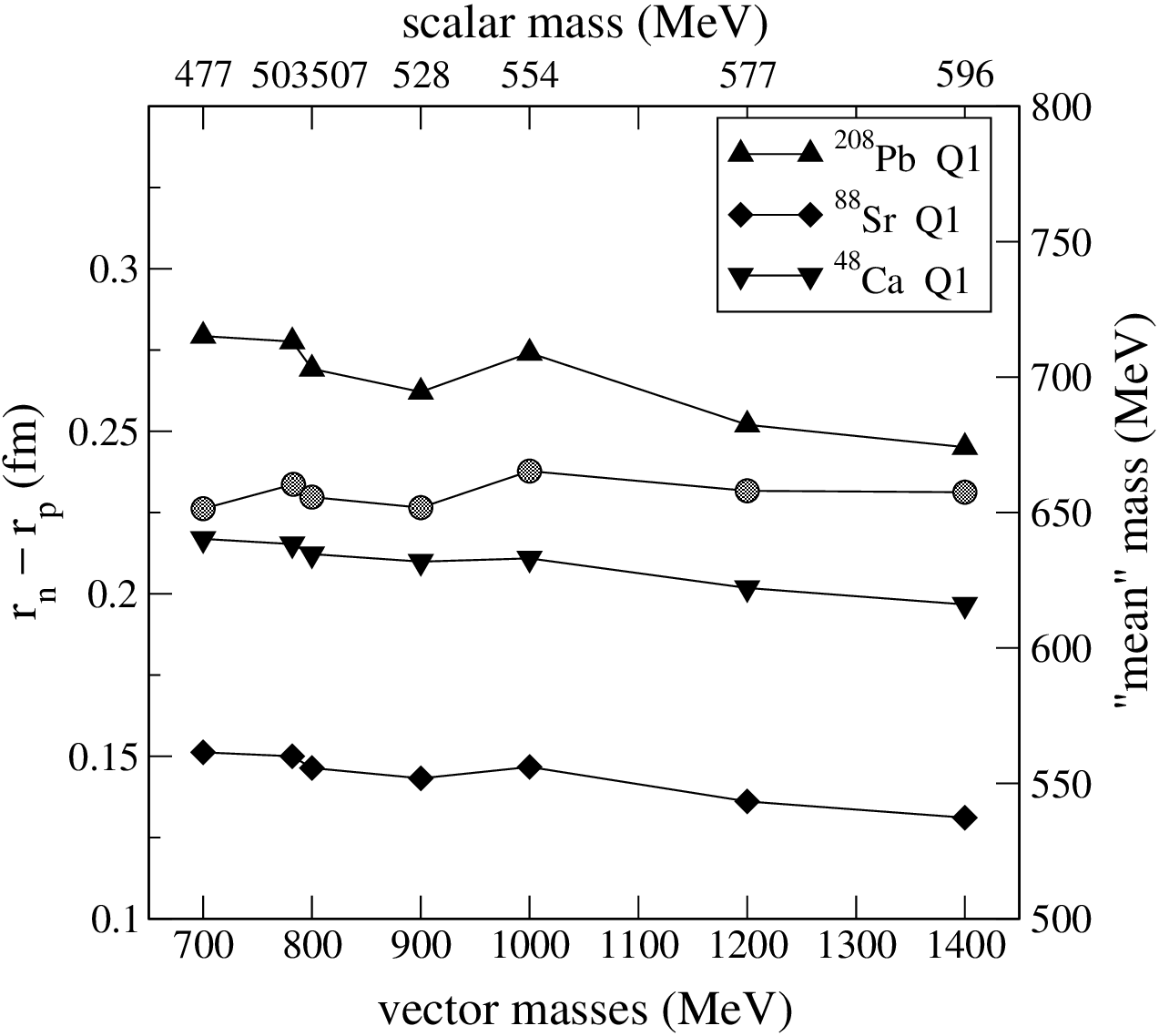}

\caption{Neutron skin thickness for three nuclei
vs.\ the isoscalar vector (bottom axis) and scalar (top axis) meson masses in
a relativistic mean-field model (Q1) from Ref.~\cite{FURNSTAHL97}.    
The circles denote the ``mean'' meson mass (right axis).}
\label{fig:rnrpmass}
\vspace*{.2in}
\includegraphics[width=4.0in,angle=0,clip=true]{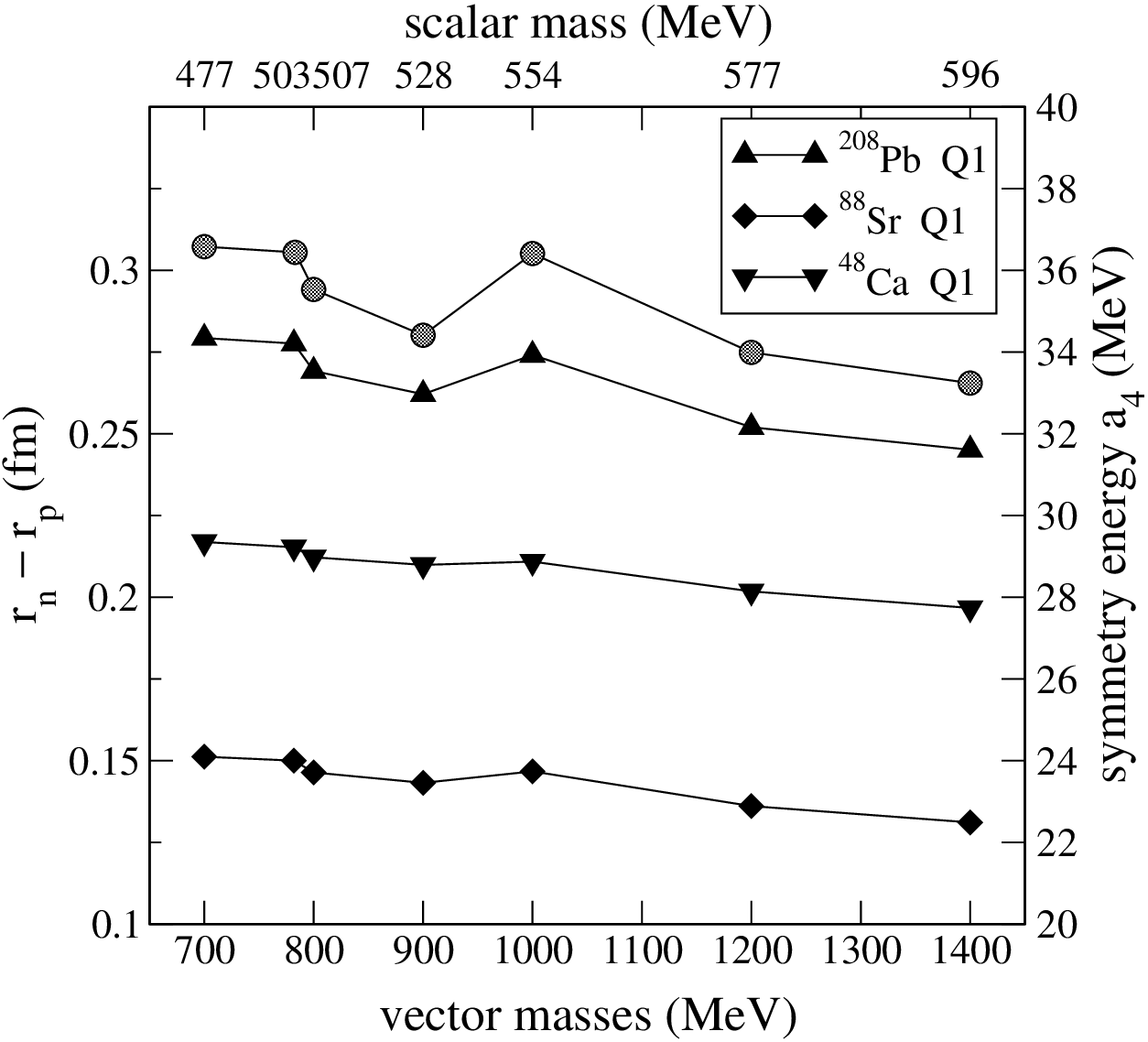}
\caption{Neutron skin thickness for three nuclei
vs.\ the isovector vector (bottom axis) and scalar (top axis) meson masses in
a relativistic mean-field model (Q1) from Ref.~\cite{FURNSTAHL97}.    
The circles denote the bulk symmetry energy $a_4$ 
(right axis) in the
adjusted parameter set.}
\label{fig:rnrpmassp}
\end{center}
\end{figure}
%%%%%%%%%%%%%%%%%%%%%%%%%%%%%%%%%%%%%%%%%%%%%%%%%%%%%%%%%%%%%%

At tree level, which corresponds to treating the  functional as
arising from the Hartree approximation, the
analysis is trivial.
The static propagator of a meson with mass $m\sim\Lambda$ can
be expanded in a Taylor series about $|{\bf q}|^2\equiv q^2=0$:
\beq 
   \frac{m^2}{q^2+m^2} = 1 - \frac{q^2}{m^2}
      + {\cal O}(q^4/m^4) 
      \approx 1 - \frac{q^2}{m^2}
      \ .
      \label{eq:prop}
\eeq
In general, there are two ingredients to the conclusion that the
higher gradient terms can be neglected in an energy functional:  
i) $q^2 \ll \Lambda^2$
for typical $q^2$
and ii) the coefficient is of order unity [as in Eq.~(\ref{eq:prop})].
A predictive power counting prescription relies on both.
For nuclei such as lead, $\langle q^2 \rangle$ 
is small in the interior and is largest
in the surface, where the scale is set by the surface thickness
$\sigma$.
Typically $1/\sigma\Lambda \approx 1/5$
\cite{FURNSTAHL99b}.
The power counting analysis in Ref.~\cite{FURNSTAHL99b} shows that 
resolving individual mass terms, which would mean determining two
terms in each isoscalar meson propagator, is not possible (that is, only one
gradient term is constrained by the fit).

We can test this conclusion directly by varying the heavy meson
masses.
In Fig.~\ref{fig:rnrpmass}, the skin thicknesses of three nuclei
are plotted against the isoscalar vector meson mass in a covariant
meson model.  As the vector mass was varied over
a factor of two, the model parameters were adjusted to achieve roughly
the same goodness of fit.  The corresponding scalar mass is shown on the
top axis (note that it varies much less than a factor of two).
While strict attention to the $\chi^2$ of the fit favors a vector meson mass
between 700 and 800\,MeV, the variation over the entire range
is remarkably small.  
This confirms that the range
of the individual mesons is not well determined by the fit to nuclei
and that a gradient expansion is more than adequate for this
observable.

The concept of optimal parameters for covariant
models was introduced in Refs.~\cite{RUSNAKTHESIS,RUSNAK97} 
and further exploited in
Ref.~\cite{FURNSTAHL99b}.
This is a reorganization of the power counting hierarchy 
by introducing linear combinations of parameters
based on the close
cancellations in covariant models between scalar and vector contributions.
By rewriting in terms of the optimal parameters, 
the number of parameters actually
determined by the data is manifested.
For example,  the combination:
\beq
  -\frac{\gs^2}{\ms^2} + \frac{\gv^2}{\mv^2} 
\eeq
is much better determined than each of these terms individually.
The choice of optimal parameters is not unique and will depend on the
quantity to be studied.
To identify appropriate optimal parameters for studying masses
(gradients) in covariant meson models,
we note that the mass terms can be rewritten:
\beq
  \frac12( \ms^2 \phi^2 - \mv^2 \Vzero^2)
    = \frac12 (\ms\phi + \mv\Vzero) (\ms\phi - \mv\Vzero)
    \equiv \frac12 \Phi_+ \Phi_- \ ,    
\eeq
which  defines the optimal combinations $\Phi_+$
and $\Phi_-$. 
Since $\Phi_+ \gg \Phi_-$ for meson models, 
if we rewrite the gradient terms using
these variables, we find
\beq
  \frac12 (\nabla \phi)^2 
     - \frac12 (\nabla V_0)^2
     = \frac18 \frac{1}{\overline m^2} (\nabla \Phi_+)^2  
     + {\cal O}(\nabla\Phi_+\cdot\nabla\Phi_-,(\nabla\Phi_-)^2)
     \ ,
\eeq
with
\beq
     \frac{1}{\overline m^2} \equiv \frac{1}{m_s^2} - \frac{1}{m_v^2}
     \ .
\eeq
This implies that the ``mean'' mass $\overline m$ is much better determined
than either of the individual meson masses.
This conclusion is verified in Fig.~\ref{fig:rnrpmass}, in which
$\overline m$ is plotted as circles (the scale is on the  right
axis).
The mass $\overline m$  varies by less than 20\,MeV over the entire range of 
variation of
the scalar and vector masses. 
Thus an (isoscalar) mass scale of about 600\,MeV underlies the model;
we identify it with the scale $\Lambda$ used in the power counting.

One might argue that the usefulness of optimal parameters imply that  
a non-covariant EFT
is more appropriate.  
In Refs.~\cite{FURNSTAHL99b} and \cite{FURNSTAHL00} this
viewpoint is challenged in detail.
Here we simply note that
the large
underlying scales characteristic of covariant models 
{\em are\/} the relevant scales in the successful power counting
prescription, and that
 the same scale $\Lambda$ is consistent
with power counting in nonrelativistic Skyrme models \cite{FURNSTAHL97c}.

If we examine Fig.~\ref{fig:rnrpmass} in more detail, we observe
a slight downward trend in the skin thickness, particularly
for lead; one might think that this is the residual effect of the 
meson range.
In fact, we can account for this trend completely by considering 
another plausible candidate for the source of variation in predicted
neutron skins:
the
symmetry energy.
We introduce the parameter 
\beq
   \alpha \equiv \frac{N-Z}{A} \ ,
\eeq
and write the semi-empirical mass formula for
the energy per particle in nuclear matter at equilibrium as
\beq
   E/A = -a_v + a_4 \alpha^2 + a_s/A^{1/3} + \ldots \ .
\eeq
In Fig.~\ref{fig:rnrpmassp}, the symmetry energy $a_4$ for each
refit model is plotted as a circle (the scale is on the right axis).  
The trend in symmetry
energy is perfectly correlated with the skin thickness.

%%%%%%%%%%%%%%%%%%%%%%%%%%%%%%%%%%%%%%%%%%%%%%%%%%%%%%%%%%
\begin{figure}[p]
\includegraphics[width=3.55in,angle=0,clip=true]{rnrp_vs_a4_sym.ps}
\caption{Neutron skin thickness in $^{208}$Pb vs.\ 
symmetry energy $a_4$ for
a single covariant meson mean-field model.  The symmetry energy is 
varied by changing the rho meson coupling constant, without refitting.
The chi-square values are shown with the data points.}
\label{fig:rnrpa4p}
\vspace*{.1in}
\includegraphics[width=3.55in,angle=-90,clip=true]{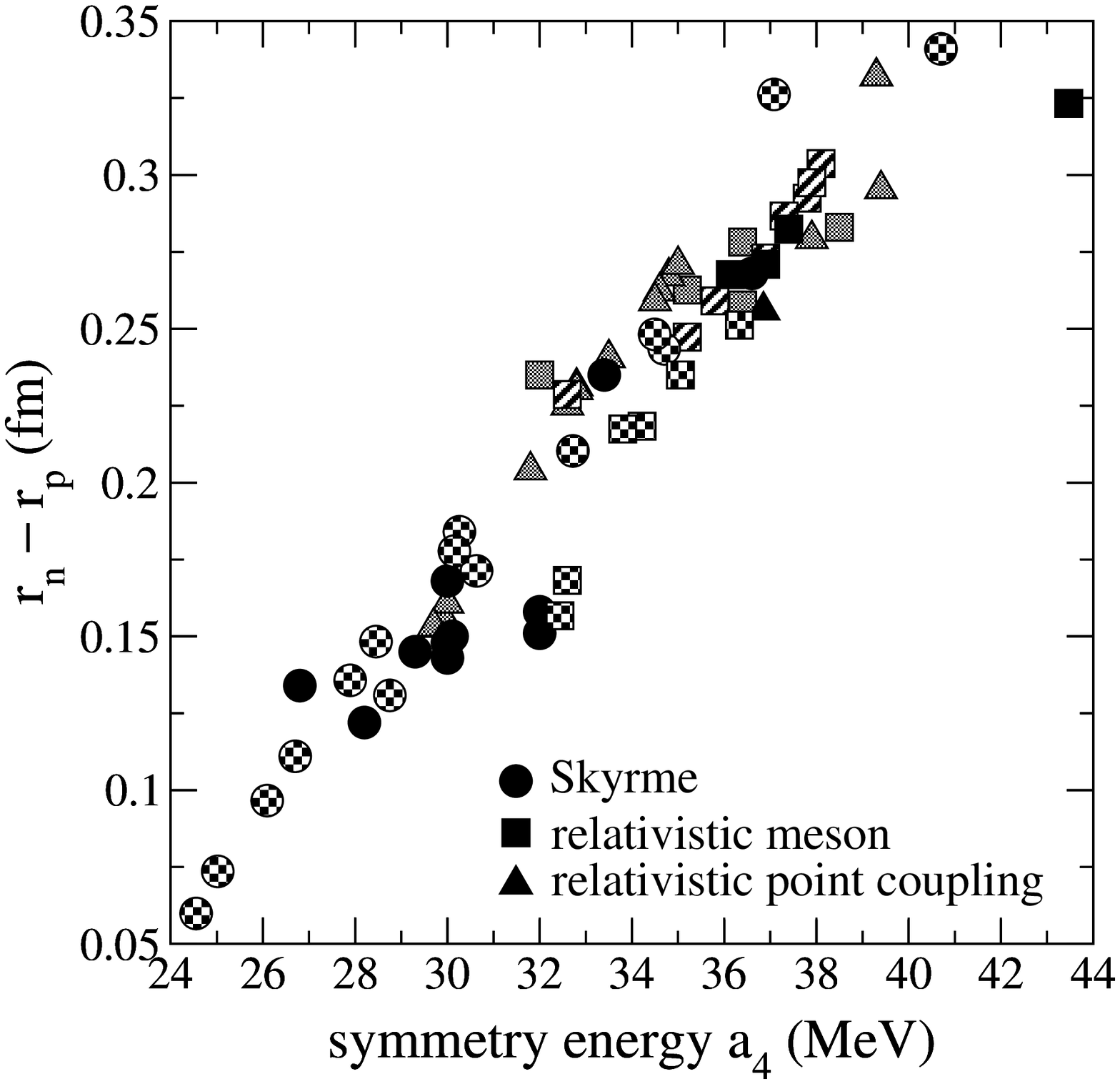}
\caption{Neutron skin thickness in $^{208}$Pb vs.\ 
symmetry energy $a_4$ for
a wide variety of mean-field models, as described in the text 
and in the caption
to Fig.~\ref{fig:rnrprn}.}
\label{fig:rnrpa4}
\end{figure}
%%%%%%%%%%%%%%%%%%%%%%%%%%%%%%%%%%%%%%%%%%%%%%%%%%%%%%%%%%%%%%

Can most of the variation
in neutron skin thickness between models be understood in terms of bulk
nuclear matter properties, or is the explicit surface symmetry energy
also important?
In Ref.~\cite{OYAMATSU}, this issue was investigated
by considering a macroscopic ``droplet-type'' model
of nuclei that divides the energy functional
for the binding energy into an infinite matter piece and
the lowest-order gradient terms.
Specifically, the binding energy $B(Z,A)$ for $Z$ protons and $A-Z$
neutrons was calculated as 
\beq
   -B(Z,A) \equiv
        \int\!\!d^3r\, E(\rho,\alpha) \rho(r)
         + \int\!\!d^3r\, F_0 ( |\nabla\rho|^2 - \beta|\nabla\rho_3|^2)
         + \mbox{coulomb} \ , 
         \label{eq:ten}
\eeq
where $\rho \equiv \rho_n(r) + \rho_p(r)$ is the baryon density,
$\rho_3 \equiv \rho_n(r) - \rho_p(r)$ and 
$\alpha(r) \equiv \rho_3(r)/\rho(r)$ is the local ratio.
Surface properties such as the neutron skin come from both density
dependence in $E(\rho,\alpha)$ and explicit contributions proportional
to $F_0$ and $\beta$. 
By fitting this model to nuclear binding energies, 
the authors of Ref.~\cite{OYAMATSU} concluded
that the explicit surface symmetry energy 
(i.e., contributions proportional to $\beta$)
was of minor importance.

%%%%%%%%%%%%%%%%%%%%%%%%%%%%%%%%%%%%%%%%%%%%%%%%%%%%%%%%%%
\begin{figure}[p]
\includegraphics[width=3.6in,angle=0,clip=true]{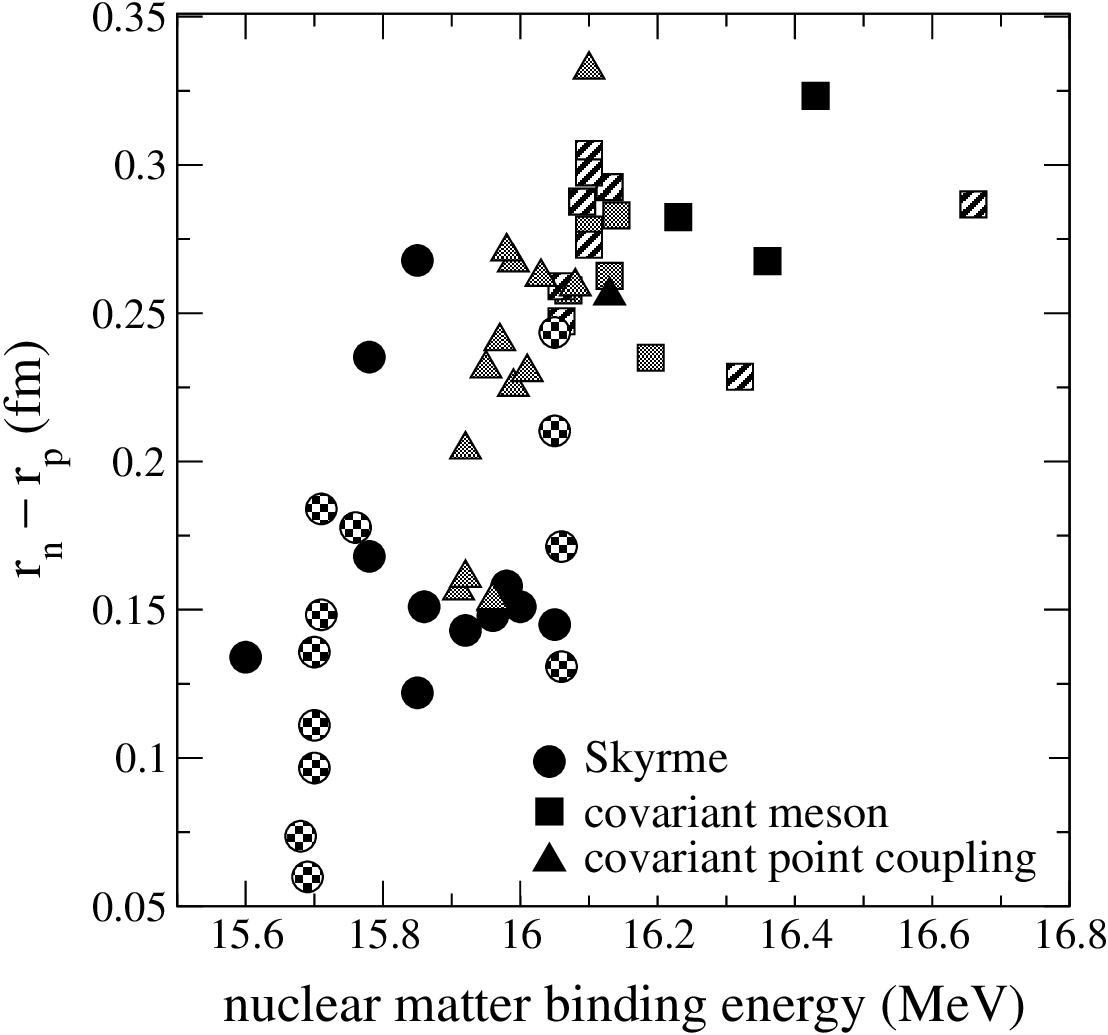}
\caption{Neutron skin thickness vs.\ nuclear matter binding
energy for 
a wide variety of mean-field models, as described in the text 
and in the caption
to Fig.~\ref{fig:rnrprn}.}
\label{fig:rnrpav}
\vspace*{.3in}
\includegraphics[width=3.6in,angle=0,clip=true]{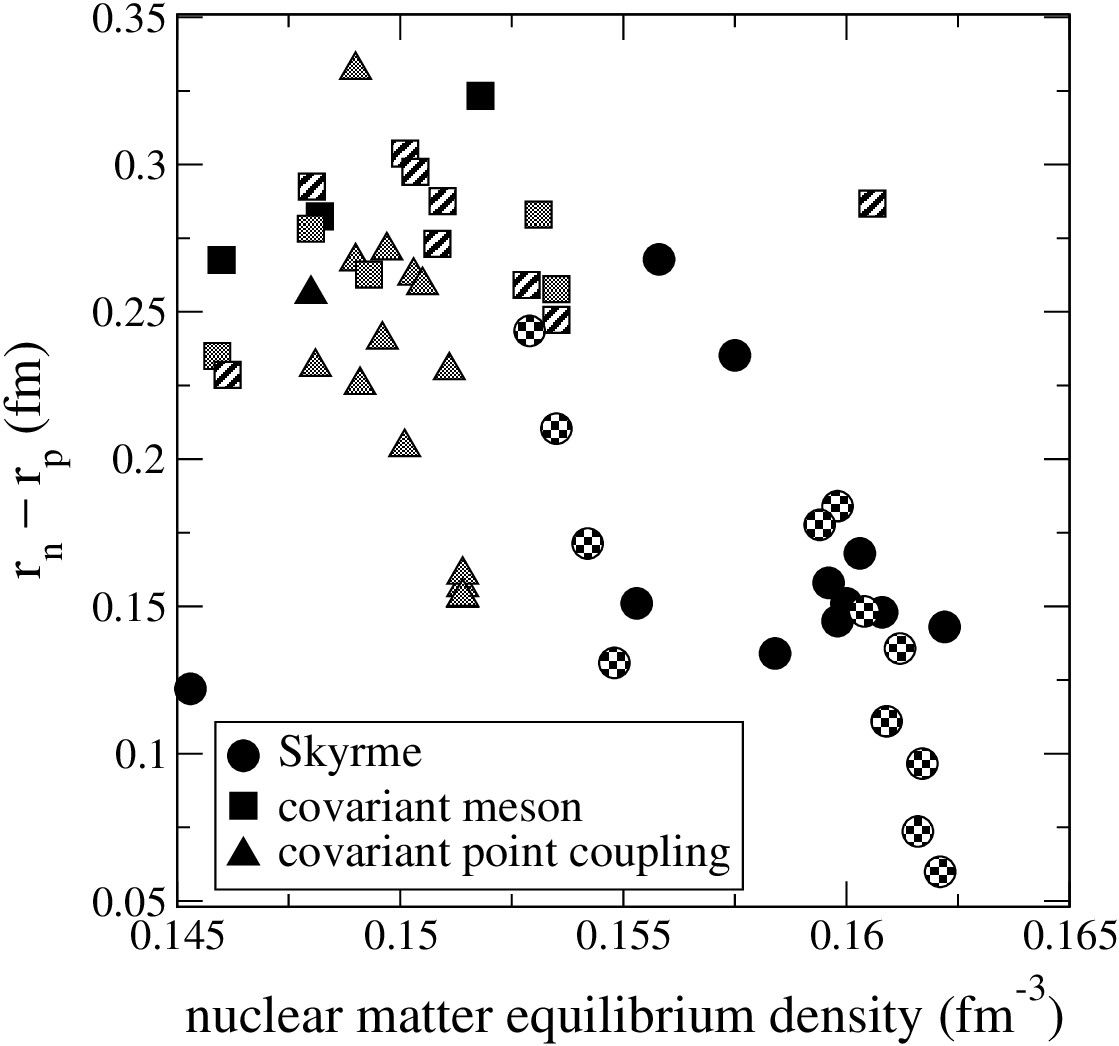}
\caption{Neutron skin thickness vs.\ nuclear matter equilibrium
density (in fm$^{-3}$) for 
a wide variety of mean-field models, as described in the text 
and in the caption
to Fig.~\ref{fig:rnrprn}.}
\label{fig:rnrprho}
\end{figure}
%%%%%%%%%%%%%%%%%%%%%%%%%%%%%%%%%%%%%%%%%%%%%%%%%%%%%%%%%%%%%%

%%%%%%%%%%%%%%%%%%%%%%%%%%%%%%%%%%%%%%%%%%%%%%%%%%%%%%%%%%
\begin{figure}[p]
\includegraphics[width=3.6in,angle=0,clip=true]{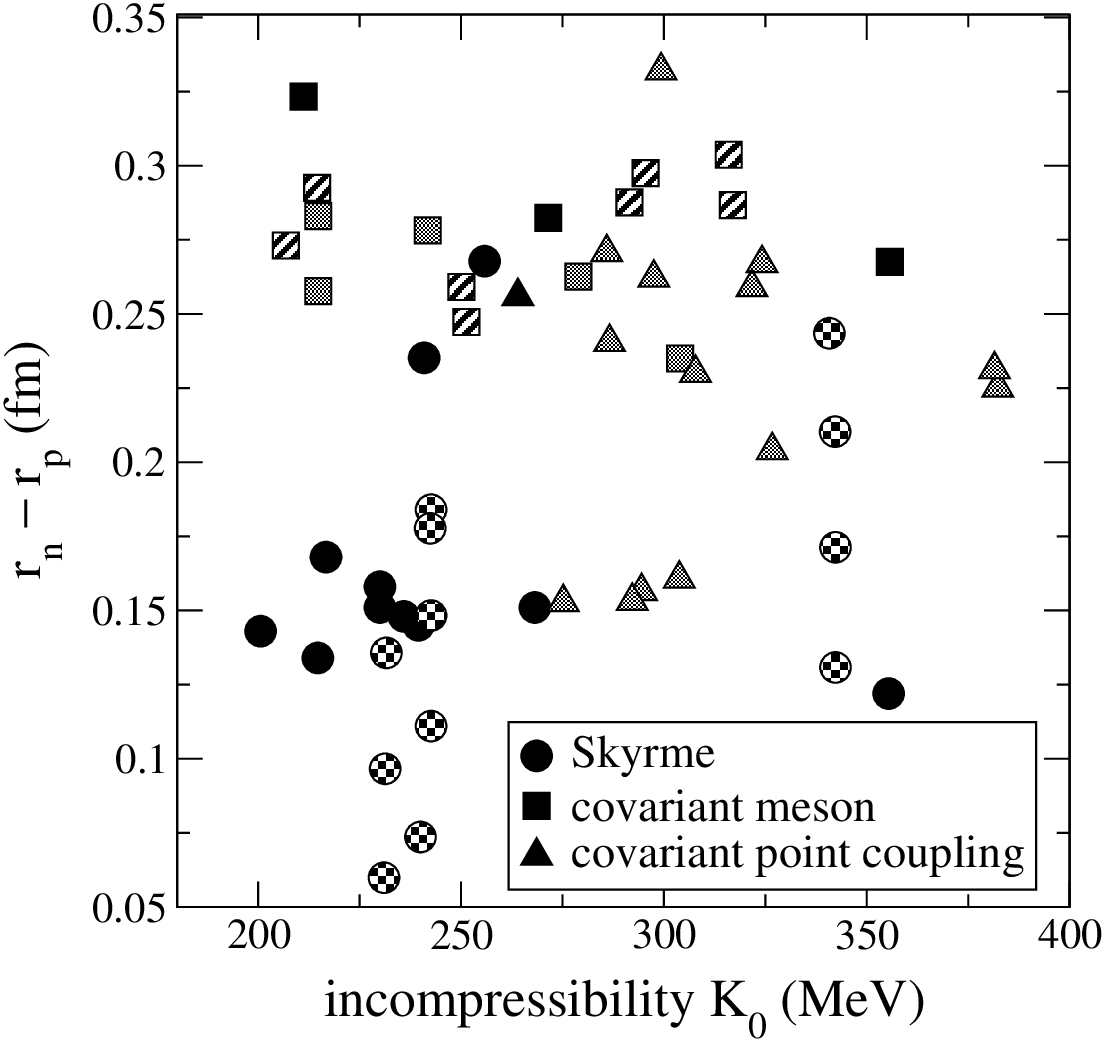}
\caption{Neutron skin thickness vs.\ nuclear matter incompressibility
for 
a wide variety of mean-field models, as described in the text 
and in the caption
to Fig.~\ref{fig:rnrprn}.}
\label{fig:rnrpK}
\vspace*{.3in}
\includegraphics[width=3.6in,angle=0,clip=true]{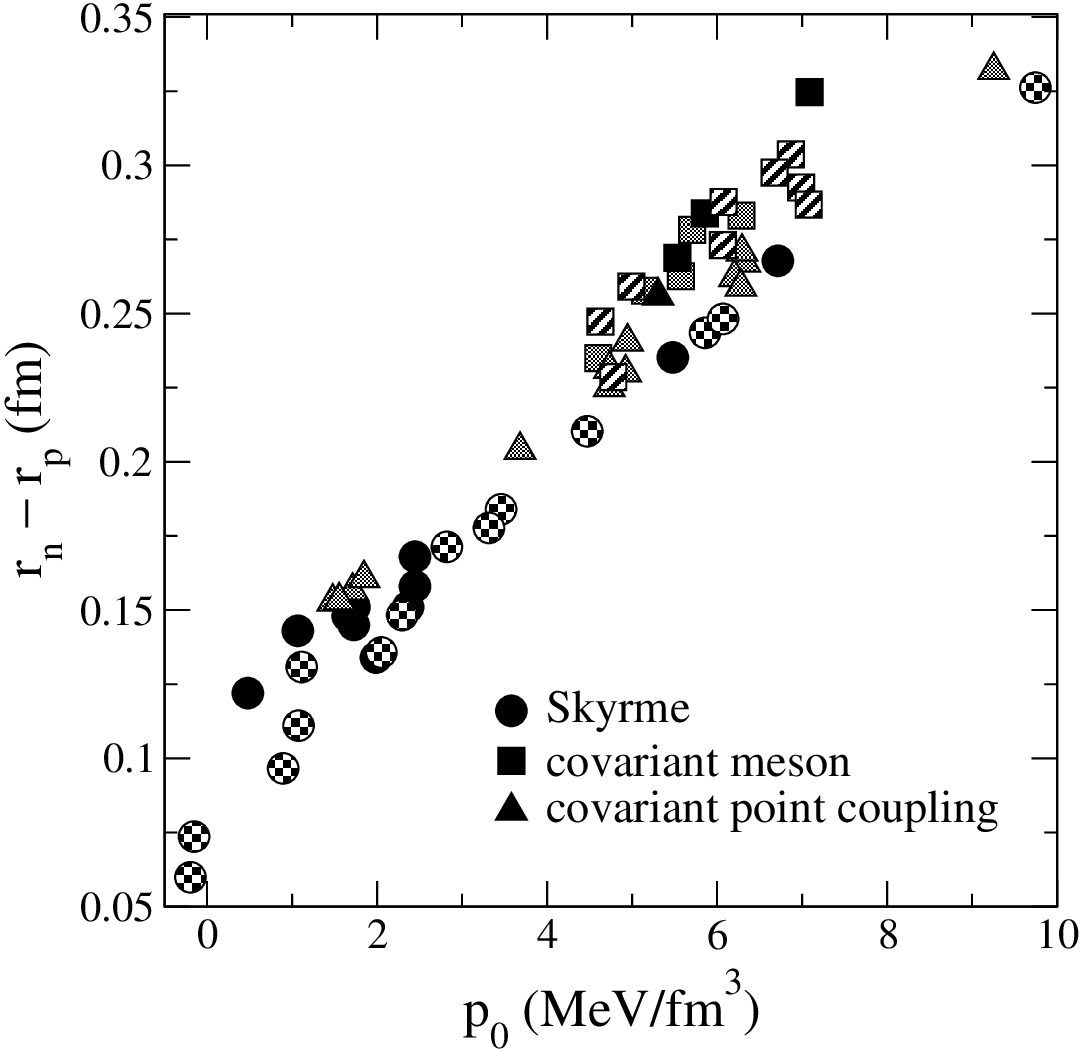}
\caption{Neutron skin thickness in $^{208}$Pb vs.\ 
linear density dependence of the symmetry energy $p_0$ for
a wide variety of mean-field models, as described in the text 
and in the caption
to Fig.~\ref{fig:rnrprn}.}
\label{fig:rnrpp0}
%
%\vspace*{.5in}
\end{figure}
%%%%%%%%%%%%%%%%%%%%%%%%%%%%%%%%%%%%%%%%%%%%%%%%%%%%%%%%%%%%%%

This model is quite compatible with the density functional
approach.
In fact,
we  have precisely the form of Eq.~(\ref{eq:ten}) with Skyrme and covariant
point-coupling energy functionals, and
for meson models one can use the meson equations
iteratively to get estimates for the parameters $F_0$ and $\beta$. 
We can estimate (and bound)
the contribution from the surface symmetry energy
by adopting {\em natural\/}
values of the associated parameters and then observing the variation
in $r_n-r_p$ as $\beta$ runs over the natural range.
The NDA estimate, which applies directly to either Skyrme or covariant
point-coupling mean-field models,
is
\beq
  F_0 ( |\nabla\rho|^2 - \beta|\nabla\rho_3|^2)
  \sim f_\pi^2\Lambda^2 \left[
  {1\over \Lambda^2}{1\over (f_\pi^2\Lambda)^2}
  \Bigl\{
    {\cal O}(1) - \frac{1}{4}\alpha^2 {\cal O}(1)
  \Bigr\}
  \right] \ .
\eeq
Numerical estimates for $\Lambda = 600\,$MeV are 
$F_0 \sim (\hbar c)^5/\fpi^2\Lambda^2 \sim 75\,\mbox{MeV--fm}^5$
and $\beta \sim {\cal O}(1/4)$, with the signs undetermined.
(The factor of 1/4 originates with the association
$\psidagger(\frac12\tau_3)\psi \leftrightarrow \frac12 \rho_3$.)
These values are consistent with all of the Skyrme and point-coupling model
fits, which is another verification of naturalness.

%%%%%%%%%%%%%%%%%%%%%%%%%%%%%%%%%%%%%%%%%%%%%%%%%%%%%%%%%%
\begin{figure}[p]
\includegraphics[width=3.5in,angle=0,clip=true]{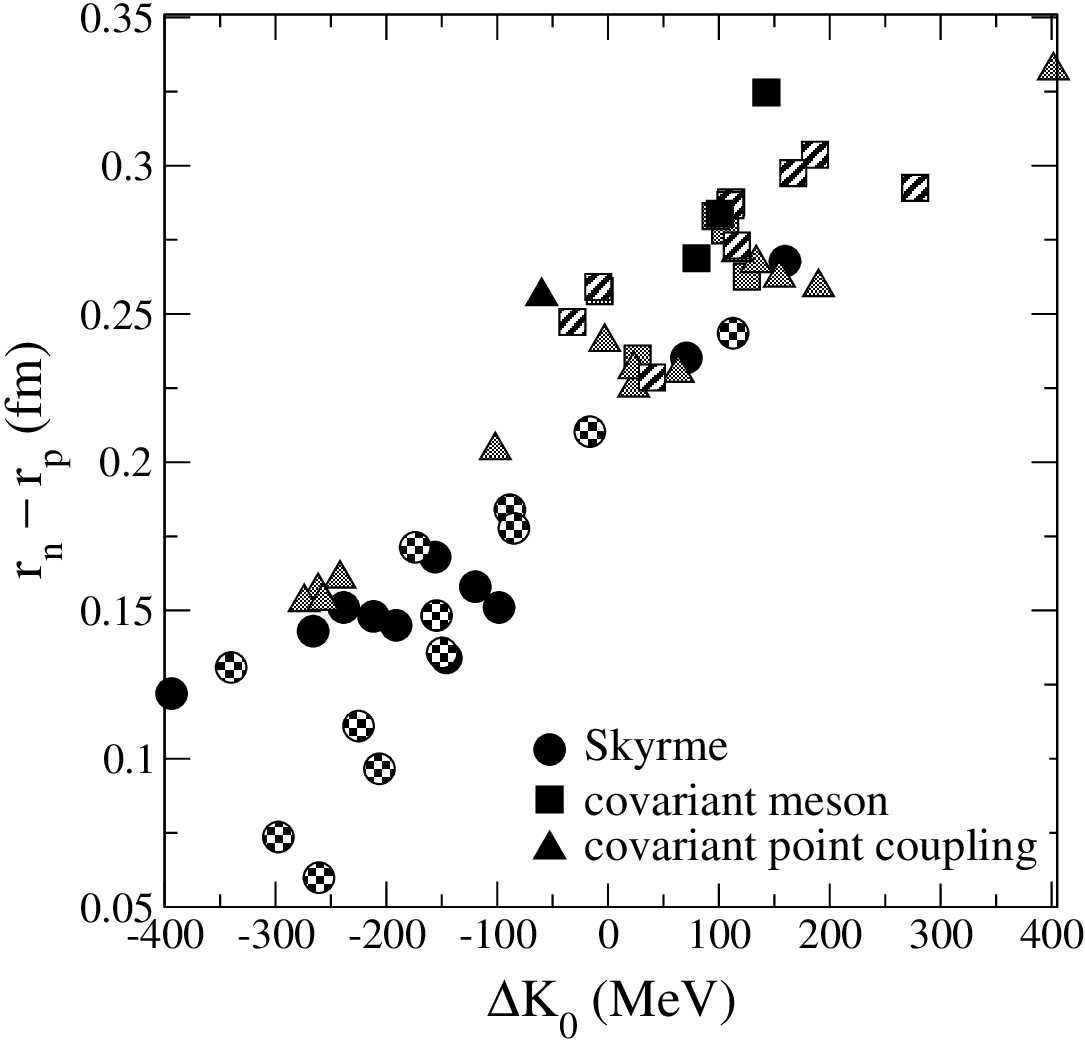}
\caption{Neutron skin thickness in $^{208}$Pb vs.\ 
$\Delta K$ [see Eq.~(\ref{eq:S2eq})] for
a wide variety of mean-field models, as described in the text 
and in the caption
to Fig.~\ref{fig:rnrprn}.}
\label{fig:rnrpDeltaK}
\vspace*{0.3in}
\includegraphics[width=3.5in,angle=0,clip=true]{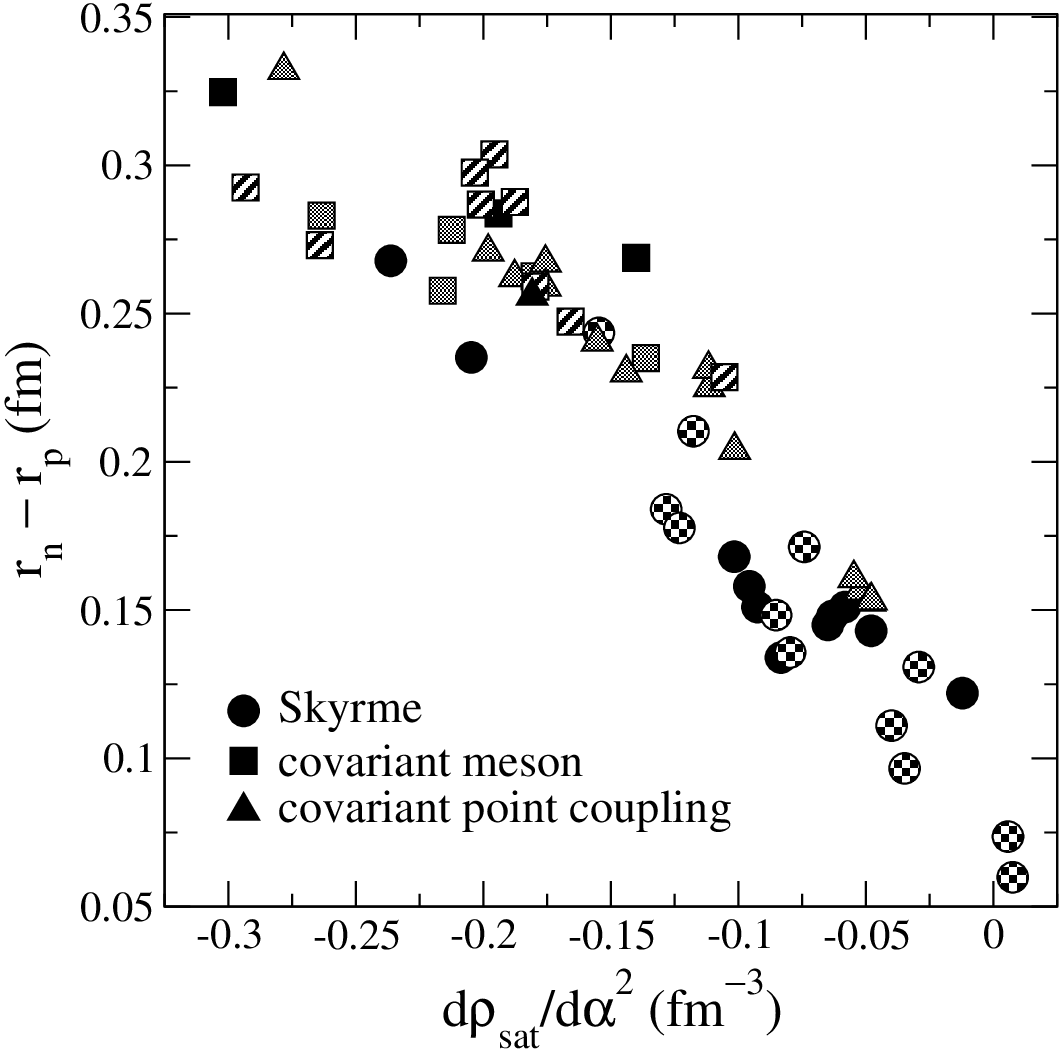}
\caption{Neutron skin thickness in $^{208}$Pb vs.\ 
$d\rho_{\rm sat}/d\alpha^2$ for
a wide variety of mean-field models, as described in the text 
and in the caption
to Fig.~\ref{fig:rnrprn}.}
\label{fig:rnrpdrho0}
\end{figure}
%%%%%%%%%%%%%%%%%%%%%%%%%%%%%%%%%%%%%%%%%%%%%%%%%%%%%%%%%%%%%%

Varying $\beta$ between $-1$ and $+1$ leads to a change in the neutron
skin thickness of less than 0.03\,fm for $^{208}$Pb,
which is small compared to the typical spread between covariant
and Skyrme models.
Thus the explicit surface dependence of the symmetry energy is not
a significant factor.%
\footnote{There is a loophole here: If $\beta$ is actually determined 
by long-range pion physics, the scale of the gradient might be
set by $m_\pi \ll \Lambda$, which would mean $\beta$ could be much larger
than unity.}
This means that we can focus our attention on nuclear matter properties.

In general,
the energy per particle of asymmetric matter
can be expanded about the equilibrium density $\rho_0$
in a Taylor series in $\rho$ and $\alpha$
(we follow the notation of Ref.~\cite{LEE98}:
\beq
  E(\rho,\alpha) 
    = E(\rho,0) 
        + S_2(\rho) \alpha^2 + S_4(\rho) \alpha^4
        + \cdots   \qquad  \alpha \equiv {N-Z \over A}
\eeq
\beq
  E(\rho,0) = - a_v + {K_0 
           \over 18 \rho_0^2} (\rho - \rho_0)^2 + \cdots
\eeq
\beq
  S_2(\rho) = a_4 + {p_0\over \rho_0^2} (\rho - \rho_0)
            + {\Delta K_0 \over 18 \rho_0^2} (\rho - \rho_0)^2 + \cdots
             \ ,
        \label{eq:S2eq}
\eeq
which defines the linear density dependence of the symmetry energy,
$p_0$, and the correction to the incompressibility, $\Delta K_0$.
These parameters and the isoscalar equilibrium parameters $a_v$,
$\rho_0$, and $K_0$ are our candidates for correlations with the
neutron skin.
The contribution of $S_4(\rho)$ for $^{208}$Pb is very small because
of the $\alpha^4$ factor and so $S_4$ is not constrained in mean-field
models, making
extrapolations to $\alpha \approx 1$ (neutron matter) quite
uncertain (although $S_2$ dominates near $\rho_0$ in any case).
We note that a study using realistic interactions found 
$S_4(\rho)$ to be unimportant even at higher densities \cite{LEE98}.

One might try to study the correlations quantitatively by varying 
appropriate parameters one by one within a single model,
as illustrated in Fig.~\ref{fig:rnrpa4p}.
The symmetry energy at equilibrium
in a single relativistic meson mean-field model
(set Q1 from Ref.~\cite{FURNSTAHL97}) 
is varied by changing the value of the rho coupling
constant $g_\rho$ [see Eq.~(\ref{eq:a4eq})].
It is clear from the figure
that the neutron skin is linearly correlated with the
symmetry energy.
But if instead we plot in Fig.~\ref{fig:rnrpa4} the skin thickness
versus $a_4$ for a wide range of models, {\em each\/} with a good fit to
nuclear properties (binding energies, charge radii, spin-orbit
splittings), we see strong correlation again but with a significantly
different slope. 
The problem is that one cannot vary one parameter independently and
expect that the fit remains good.
The goodness-of-fit value (labeled $\chi^2$) for each set
is shown in Fig.~\ref{fig:rnrpa4p}, with a value of 100 indicating an
acceptable fit.
This shows that the quality deteriorates rapidly unless the other
parameters are allowed to change.
A more sophisticated approach to the correlation analysis 
during the fitting process would
be very helpful in future investigations.

In Figs.~\ref{fig:rnrpav}, \ref{fig:rnrprho}, and \ref{fig:rnrpK},
the correlations of the skin thickness with the symmetric nuclear matter
($\alpha=0$) binding energy, equilibrium density, and
incompressibility are shown.  The coarse observation is that there are
no dramatic correlations between these quantities individually and the
skin thickness.  At a more fine-grained  level one can note:
\begin{itemize}
 \item The binding energy of Skyrme models is systematically lower
 than for covariant models, although the difference is small and there
 is significant overlap.
 \item The equilibrium density of Skyrme models is systematically
 larger than for covariant models.  This has long been observed. 
 There is a very rough correlation with skin thickness.
 \item The incompressibility seems largely uncorrelated with the skin
 thickness and there are no clear patterns between different types of
 models.
\end{itemize}
Overall, we find no signature that the physics determining these
quantities directly determines the skin thickness.
Secondary correlations are expected simply from the requirement that a
good fit is always obtained.

In Figs.~\ref{fig:rnrpa4}, \ref{fig:rnrpp0}, and \ref{fig:rnrpDeltaK},
we show the correlations between skin thickness and the individual
parameters describing the symmetry energy $S_2(\rho)$.
In each case the correlation is very strong and approximately linear.
The spread for $p_0$, in particular, is remarkably small and there is
no significant separation of the different models:  to good
approximation all covariant and Skyrme models 
lie on the same line.  
A clear corollary to the correlations of $a_4$, $p_0$, and $\Delta
K_0$ to the neutron skin thickness is that
these parameters
are also highly correlated with each other.

%\section{Density Dependence of the Symmetry Energy}\label{sec:density}

\section{Discussion}\label{sec:discussion}

Qualitatively, the neutron skin is determined from the energy balance sought
for $N \neq Z$ between the extremes of equal proton and neutron radii
but different densities, and equal densities but a sizable neutron
skin region.
Thus,
the cost in energy is determined in large part by the density dependence of the 
symmetry energy, $S_2(\rho)$ \cite{OYAMATSU}.
In addition, the Coulomb repulsion favors larger proton radii,
which reduces the size of the neutron skin in lead approximately 0.1\,fm
relative to the radii with  Coulomb turned off,
with only a small residual dependence on the symmetry energy.
(This result was obtained by comparing radii with and without the
Coulomb interaction for each model.)
A simple density-expansion
analysis based on these ideas made by Oyamatsu et al.\
\cite{OYAMATSU} identifies as a key parameter the shift in saturation point
with the asymmetry, $d\rho_{\rm sat}/d\alpha^2 = -9p_0/K_0$.
The correlation of the neutron skin
with $d\rho_{\rm sat}/d\alpha^2$ is shown in
Fig.~\ref{fig:rnrpdrho0} and is strong, 
although there is greater spread than with
$p_0$ alone. 

\begin{figure}[t]
\includegraphics[width=3.0in,angle=0,clip=true]{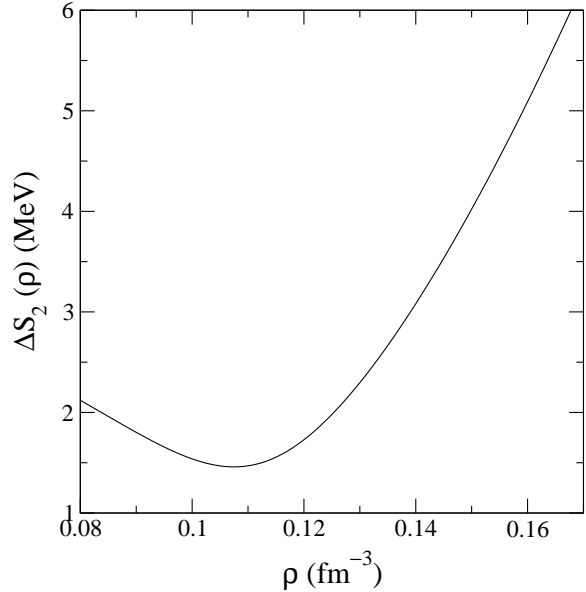}
\caption{The standard deviation of $S_2(\rho)$ values for all of
the mean-field models as a function of $\rho$.}
\label{fig:S2sdrho}
\end{figure}

An explanation of the strong correlation in mean-field models
between the parameters of
$S_2(\rho)$ has been offered by Horowitz \cite{HOROWITZPC}.
There is not enough resolution in the binding energy systematics 
of finite nuclei to fix
$S_2(\rho)$ for a range of densities.  In fact, as discussed earlier,
only one
quantity is resolved, so one determines the symmetry energy only at an average
density $\langle \rho \rangle$ for nuclei.
(For lead we find $\langle \rho \rangle \approx
0.11\,\mbox{fm}^{-3}$.)
Consequently there are many combinations of $a_4$, $p_0$, and $\Delta
K_0$ (and $\rho_0$) that reproduce this value.  This conjecture is
supported by the observation that the spread of
$S_2(\langle\rho\rangle)$ values among the models used
in the figures  is several times smaller than the spread of
 $S_2(\rho_0)$ values, as shown in Fig.~\ref{fig:S2sdrho}.

%%%%%%%%%%%%%%%%%%%%%%%%%%%%%%%%%%%%%%%%%%%%%%%%%%%%%%%%%%
%%%%%%%%%%%%%%%%%%%%%%%%%%%%%%%%%%%%%%%%%%%%%%%%%%%%%%%%%%%%%%

The standard parameterization of energy functionals for covariant
meson models
leads to an enhanced symmetry energy because the kinetic energy
piece is large (since
$\Mstar$ is reduced from $M$) and the contribution from
the $\rho$ meson mean field is positive definite.  
Specifically, $a_4$ is given by
 \beq
   a_4 =  \frac{1}{6}\frac{\kfermi^2}{(\kfermi^2+\Mstar{}^2)^{1/2}}
     +  \frac{g_\rho^2}{8 m_\rho^2}\rho_0
     \ ,
     \label{eq:a4eq}
 \eeq
where $g_\rho$ is the coupling, $m_\rho$ is the $\rho$ mass parameter,
$\kfermi$ is the Fermi momentum, and $\Mstar$ is the effective nucleon
mass \cite{SEROT86}. 
In addition,
$p_0$ scales
with $a_4$ (e.g., the $\rho$ contribution to $p_0$ is
$g_\rho^2\rho_0^2/8m_\rho^2$).
If fits to nuclei determine the symmetry energy at some averaged
density less than equilibrium density $S_2(\langle\rho\rangle)$, 
this implies that both
$a_4$ and $p_0$ will be large (i.e., a large density dependence).  
In fact, typical values
for $p_0$ are around 6\,MeV/fm$^3$, compared to about 
2\,MeV/fm${^3}$ for most standard Skyrme models, while
the corresponding values for $a_4$ are around 36\,MeV and 30\,MeV.

Given this association, we can ask:
what should $p_0$ be? 
Engvik et al.\ compared calculations of asymmetric matter
using the best-available nonrelativistic potentials and many-body
calculations \cite{ENGVIK97}.
From the slopes of the curves given in Ref.~\cite{ENGVIK97},
one can determine ${p_0}$ (up to a factor
of ${\rho_0^2}$).  One finds $p_0 \approx 2\, \mbox{MeV/fm}^3$ with only
about a 10\% spread.  [Note:  there is some ambiguity in the
choice of  ${\rho_0^2}$ to compare with
mean-field models; if a larger value associated with some Skyrme
models was used, then ${p_0}$ would be 
10--30\% higher.]
Two Dirac-Brueckner studies
of asymmetric nuclear matter \cite{SUMIYOSHI95,LEE98} yield 
somewhat higher numbers, $p_0 \approx 3\,\mbox{MeV/fm}^3$.
However, both sets of values
are significantly less than those from
the standard covariant mean-field meson models
(i.e., the solid black squares in Fig.~\ref{fig:rnrpp0}). 
Generalized meson models can be forced to have lower $p_0$
(by adding a low neutron radius to the fit observables) but
sets with $p_0$ lower than 5\,MeV/fm$^{3}$ are only achieved if the
additional isovector parameter ($\eta_\rho$ from
Ref.~\cite{FURNSTAHL97}) becomes highly unnatural.

Generalized point-coupling models \cite{RUSNAK97}, which feature higher-order
terms not included in conventional covariant models, were found
with lower values of $p_0$ and good fits (see the shaded triangles in
Fig.~\ref{fig:rnrpp0}).
However, this lower value arises  from cancellations 
between two orders in 
the density expansion for the isovector contribution, which again violates
the principle of naturalness.
In particular, the contribution to $p_0$ analogous to the $\rho$ meson
contribution is proportional to $\wt\zeta_2 +
\frac32\wt\eta_\rho\wt\rho_0$,
where $\wt\rho_0 \approx 1/5$.
By choosing 
${\wt\eta_\rho \approx -10/3}$, one can cancel the leading
contribution for natural $\wt\zeta_2 \approx 1$.  
In fact, one of the sets has $\wt\eta_\rho = -3.245$.
We do not have a physical
reason for such a fine tuning between different orders
and conjecture that it may reflect instead limitations in the
isovector contribution to the energy functional.   

The conclusion is therefore that we need to revisit isovector
physics in covariant models.
There are many possible deficiencies to explore:
\begin{itemize}
  \item The functional is incomplete.
  For example, isovector scalar contributions (the ``$\delta$'' meson)
  may be important for isovector properties.
  This possibility has been recently considered in Ref.~\cite{LIU01}.
  \item The functional is truncated too severely.  The higher-order
  terms discussed above are an example of what could be added.
  \item Long-range physics must be explicitly included, as expected in
  an EFT treatment.  One natural class of contributions is from the pion.
  As noted above, explicit pions have not been {\em necessary\/}
  because
  the isoscalar part is shorter-ranged (``$\sigma$'' physics)
  and the isovector part is poorly resolved. 
  However, power counting implies that pion contributions should be
  included.
  In addition, 
  many-body long-range correlations (e.g., RPA correlations) may
  be needed. 
  \item The one-loop form of the functional is not sufficiently general.
  For example, one will obtain
  nonanalytic terms in the density from the long-range contributions.  
\end{itemize}
We seek a systematic approach to incorporating this physics into our
descriptions of medium and heavy nuclei.
Furthermore, a connection to underlying forces is desirable to help
constrain the functional.

A key question is whether mean-field models can be made more systematic
without drastically increasing the computational burden.
A great appeal of mean-field models is the ease with which they can
be used to calculate medium to heavy nuclei.  The basic 
calculational procedure is the same
for a covariant Hartree or nonrelativistic Skyrme interaction.
The models can all be formulated in terms of an energy
functional, whose minimization with respect to single-particle orbitals
yields a single-particle ``potential''
functional. 
This potential is used in turn to calculate the orbitals. 

The generic procedure is quite simple.
To solve for the ground-state bulk properties of a nucleus with
$A = Z + N$ nucleons,
\begin{enumerate}
  \item guess a set of 
    initial density/current profiles $\{\rho(\xvec)\}$ (this notation
    denotes the
    baryon and isospin densities, plus any other spin or scalar
    densities);  most can be set to zero or initialized with a Fermi shape
    with appropriate parameters;
  \item evaluate
  a functional of the  $\{\rho(\xvec)\}$, yielding 
  a {\em local\/} single-particle potential $\Vs(\xvec)$;
  \item solve the Dirac or Schr\"odinger equation for the lowest
  $A$ eigenvalues and eigenfunctions $\{\epsilon_\alpha,\psi_\alpha\}$:
 \beq
  \bigl[ -i\bm{\alpha}\cdot\bm{\nabla} + \beta M
     + \beta \Vs(\xvec)
  \bigr]\, \psi_\alpha(\xvec) = \epsilon_\alpha \psi_\alpha(\xvec) \ , 
 \eeq
 or
 \beq
  \biggl[ -\frac{{\nabla}^2}{2M}  +  \Vs(\xvec)
  \biggr]\, \psi_\alpha(\xvec) = \epsilon_\alpha \psi_\alpha(\xvec) \ , 
 \eeq
 \item compute new densities, e.g., 
  \beq
    \rho(\xvec) = \sum_{\alpha=1}^{A} |\psi_\alpha(\xvec)|^2 \ ,
  \eeq
  and other observables (which are functionals of 
  $\{\epsilon_\alpha,\psi_\alpha\}$);
  \item repeat steps 2.\--4.\ until the changes from iteration
  to iteration are acceptably small, i.e., until the solution
  is self-consistent.
\end{enumerate}

This procedure is  straightforward to implement numerically.
Dealing with deformed nuclei and open shells introduces some
complications (e.g., pairing), but these do not qualitatively complicate
the calculations.
There are two key reasons for the simplicity:
\begin{itemize}
  \item the energy functional is {\em universal\/}, in the sense
  that the same functional is used for all nuclei, with the same
  set of parameters; 
  \item the single-particle functional, which is derived from the
  functional derivative of the energy functional with respect to
  $\rho(\xvec)$ is {\em local\/}
  (at least when pairing is treated most simply) {\em and\/} is easily
  evaluated in terms of the eigenvalues and eigenfunctions.
\end{itemize}
Once the functional is established, each subsequent
finite nucleus calculation is almost
trivial.  Thus it is certainly desirable to maintain this mean-field 
calculational procedure.

In Ref.~\cite{AKMAL98}, it is argued that the mean-field approximation
for meson-ranged interactions cannot be valid at ordinary densities, 
because the inverse
Compton wavelengths $\mu$ of the mesons times the mean interparticle
spacing is much less than unity.  The authors go on to state that
``The RMF approximation can be based on effective values of the coupling
constants that take into account the correlation effects.  However,
these coupling constants then have a density dependence, and a
microscopic theory is needed to calculate them.''
The second objection is misleading, since
density dependent couplings are not necessary; all of the density
dependence can be incorporated in the functional, if made sufficiently
general.
The correspondence between
models with explicitly density dependent couplings and more
general functionals can be made through field redefinitions
\cite{FURNSTAHL00}.
Matching to a microscopic theory may be a desirable way to determine
the parameters, but it is not necessary.  Instead, one can match to
finite density observables, which is especially needed for calculating
heavy nuclei.

To address the first objection,
we propose a merger of Density Functional Theory (DFT) and
Effective Field Theory (EFT).  
Kohn-Sham DFT is a framework
that looks just like the simple relativistic Hartree calculations
that are so well adapted to calculating finite systems, but one which can
include all short-range and long-range correlations
\cite{ARGAMAN00}.  
That is, the solution framework described above can accommodate
the most general description of the ground state within the context
of density functional theory, and the Kohn-Sham scheme in particular.
The implication is that conventional mean-field approaches provide reliable
descriptions of bulk properties not because the mean-field
approximation is good but because they accurately approximate exact
density functionals.
This claim is supported by the work of Hu, who showed that
Dirac-Brueckner-Hartree-Fock 
at the two hole-line level, when matched to a generalized
EFT-based functional, is reproduced with natural coefficients \cite{HU99}.

Therefore,
we interpret the ``mean-field'' functionals as just  particular
generalized gradient approximations to complete 
Kohn-Sham density functionals \cite{PERDEW96,PERDEW99}.
(Skyrme may be considered more general in this regard.)
Rather than starting from the underlying free-space interaction,
one expands the functional as a series expansion in density and
momentum, and then fits directly to experimental data.
This approach can miss (long-range) non-analytic or nonlocal
contributions, such as explicit exchange (``Fock'') terms
(e.g., from pions).
The role of the EFT framework is to add these systematically while
maintaining symmetries and conservation laws.

From a Green's function perspective, the mean-field models 
follow from the Hartree (relativistic) or Hartree-Fock (nonrelativistic)
approximation to the nucleon self-energy, with
an effective interaction (e.g., an approximation to the G-matrix).
In practice,
the nonrelativistic Skyrme interaction uses zero-range (``point couplings''),
which means that the energy functional has the same
form as for a Hartree calculation
(one can make a Fierz rearrangement of the interaction to derive the
 correspondence).
The Hartree self-energy in coordinate-space is static and local
(i.e., a function of $\xvec$ only).  

In contrast, the general self-energy is non-local in space and is
frequency dependent.  
This would seem to imply that the correlation effects that go beyond
the Hartree approximation could never be systematically included
in the calculational
framework described above; one could only incorporate averaged effects
using effective interactions.
Since the simplicity of the mean-field approach depends heavily
on the local, static nature of the single-particle potential, it would
appear that systematic improvement is not possible.
The answer is that we don't need to calculate the general self-energy.
Instead we use
the Kohn-Sham DFT potential $V_s({\bf x})$, which is static and local,
even though it can include all correlations. 

There is a danger that evaluating the single-particle functional 
to find $V_s(\xvec)$ will become too great a
computational burden when going beyond the current mean-field models.  
So developing reliable gradient expansions will
be essential.
The surprising experience from Coulomb systems that LDA plus
generalized gradient expansions work extremely well needs to be
understood in this context.
In particular, it does not rule out long-range physics being
incorporated in the same form as
existing functionals, although contributions nonanalytic in the
densities should be expected.  For example, the exchange contribution
from a long-range (i.e., effectively massless) particle will have a
$\rho^{4/3}$ contribution to the energy density.   
Thus, an EFT treatment of the pion expanded
around the chiral limit should contribute a term like this.
Work is in progress to cast Kohn-Sham DFT into an effective
action formalism for composite operators \cite{PUGLIA}.

We could imagine constructing covariant models more like Skyrme models,
by including terms like 
$\rho_s^\delta$ and $\rho_B^\gamma$, with fractional $\delta$ and
$\gamma$ determined phenomenologically.
But the  construction from a Lagrangian has been an important guide 
to maintaining covariance and building a conserving approximation.  
In addition, the connection to free space scattering is much closer than
in the nonrelativistic case since self-energies {\em are\/} dominated by
mean-field components (Hartree dominance), so corrections are smaller.
Therefore, we advocate deriving a 
DFT expansion from an EFT Lagrangian, 
or at least to establish constraints on the analytic structure.

Finally, we return to the issue of point nuclear densities.
Single-nucleon structure will automatically be included in energy
functionals fit to data.  Since
point proton and neutron radii are not observables (no probe couples
to them) they are at best auxiliary quantities.  
This is not a major issue for properties such as the
charge density alone, since the nuclear structure is mostly isoscalar
while the single-nucleon form factors are mostly isovector and so
 factorization is a good starting point.  But considering
$r_n-r_p$ heightens
sensitivity to the isovector form factor.  In the future, one should
actually couple to electroweak currents to avoid the model dependence
inherent in defining the point densities.  
One of the strengths of an EFT is knowing how to couple to external
currents, which may involve identifying new constants (low-energy
constants).

\section{Summary}\label{sec:summary}

We have examined the predictions by mean-field models
of neutron skin thicknesses, which are highly correlated with the
neutron radius.
We find that the skin thickness
is closely related to the density dependence of the symmetry
energy, independent of the type of mean field model.
In particular, the linear density coefficient $p_0$ exhibits a remarkably good
linear correspondence.
(However, the strong correlations between parameters preclude
isolating the effects of a single parameter.) 
This implies that it is a general feature of the energy functional.
Our conclusion is in accord with other analyses in the recent
literature \cite{OYAMATSU,BROWN00,BROWN01}.

An accurate extraction of the radius from experiment, for even a
single nucleus, would provide a new and valuable constraint on the
energy functional.
Current analyses of proton scattering are consistent with a skin
thickness in lead of order $0.1$--$0.2\,$fm \cite{KARATAGLIDIS01,BUNNY},
which would imply $1\,\mbox{MeV/fm}^3 < p_0 < 4\,\mbox{MeV/fm}^3$,
and recent analyses of antiprotonic atom data also lie in that
range \cite{TRZCINSKA01}.
Calculations of uniform matter with realistic potentials show a
narrower range of $2\,\mbox{MeV/fm}^3 < p_0 < 3\,\mbox{MeV/fm}^3$,
with Dirac-Brueckner calculations at the high end and nonrelativistic
calculations at the low end.
If these conclusions are validated, it means that the uncertainty in neutron
radii may be much smaller than implied by comparisons of the current ``best fit''
calculations.

However, the commonly used
mean-field functionals are not  equally flexible in accommodating
different values of $p_0$.  
In particular, covariant meson models have significantly larger values
of $p_0$ than found in most Skyrme parametrizations, such as
the recent set of Brown \cite{BROWN98}.
Calculations of uniform matter with realistic potentials
imply that $p_0$ is close to that of nonrelativistic Skyrme models
while it is significantly overestimated in the most widely used 
relativistic mean-field
models.  It is not surprising that this does not preclude a good
fit 
to the best measured properties of
finite nuclei, given the limited constraint from these properties on
purely isovector parameters.
It is not clear whether the tendency for
Skyrme models to have lower $p_0$ is a consequence of the form of
truncation or is slightly favored by the fits to other observables. 
In Ref.~\cite{BROWN98}, additional information, which corresponds to
fixing $p_0$, was used to constrain one of the isovector parameters.

Such comparisons between nonrelativistic and covariant models
might be used to argue that one approach is more
``correct'' than another.  
However, the EFT perspective changes the context of the discussion
from correctness to efficiency and the effects of truncation.
The most commonly used covariant parametrization appears to be
deficient because it does not allow lower $p_0$ with natural
parameters, but the conclusion
should be that the functional must be improved.
Power counting suggests several avenues:  a more complete set of 
operators at the lower orders (e.g., add an isovector scalar meson or density),
higher order contributions (density corrections to rho meson exchange), pion
and other long-range contributions.
Additional isovector observables, such as binding energies of more
asymmetric nuclei and observables from collective excitations will be
needed to constrain new parameters. 
The question is: how can we incorporate this physics to
improve the covariant functionals (and
nonrelativistic functionals) systematically? 
We propose developing density functional theory in an EFT framework,
using effective action techniques \cite{PUGLIA}.

\vspace*{-2pt}

\acknowledgments

We thank B.~Clark, H.-W.\ Hammer, and B.~Serot  for useful comments.
This work was supported in part by the National Science Foundation
under Grant Nos.\  PHY--9800964  and PHY-0098645.


\begin{thebibliography}{99}

% New Skyrme interaction for normal and exotic nuclei
\bibitem{BROWN98}B.~A.\ Brown, Phys.\ Rev. C {\bf 58} (1998) 220,
  and references therein.
  
% Ring review article
\bibitem{RING96}P.~Ring, Prog.\ Part.\ Nucl.\ Phys.\ {\bf 37} (1996)
  193, and references therein.  
%

% Parity violating measurements of neutron densities
\bibitem{HOROWITZ01}C.~J. Horowitz, S.~J. Pollock, P.~A. Souder, 
   and R. Michaels, Phys.\ Rev.\ C {\bf 63} (2001) 025501.


% Neutron star structure and the neutron radius of pb-208
\bibitem{HOROWITZ00}C.~J.\ Horowitz and J. Piekarewicz,  
    Phys.\ Rev.\ Lett.\ {\bf 86} (2001) 5647.

% Nuclear Structure and Astrophysics Town Meeting White paper
\bibitem{WP2000}Report of the ``Nuclear Structure and Astrophysics
  Town Meeting,'' Oakland, CA, November, 2000,
  http://snohp1.lbl.gov/$\sim$lpr2000/oakland\_wp.pdf.

% Realistic models of pion exchange three-nucleon interactions
%\bibitem{PIEPER01}S.~C.\ Pieper, V.~R.\ Pandharipande, R.~B.\ Wiringa,
%  and J.~Carlson, Phys.\ Rev.\ C {\bf 64} (2001) 014001.

% The equation of state of nucleon matter and neutron star structure
\bibitem{AKMAL98}A. Akmal, V.~R.\ Pandharipande, and D.~G.\ Ravenhall,
  Phys.\ Rev.\ C {\bf 58} (1998) 1804.

% Density Functional Theory
\bibitem{KOHN65} W. Kohn and L. J. Sham, Phys.\ Rev.\  {\bf A140} (1965) 1133.
%
\bibitem{DREIZLER90}R. M. Dreizler and E. K. U. Gross,
      {\it Density Functional Theory} (Springer, Berlin, 1990).
%
\bibitem{ARGAMAN00}N. Argaman and G. Makov, Am.\ J.\ Phys.\ {\bf 68}
   (2000) 69.	   

% Some recent developments in the mean-field description of low-energy
%  nuclear physics
\bibitem{BRACK85}M.~Brack, Helv.\ Phys.\ Acta {\bf 58} (1985) 715.

% Density functional approach to quantum hadrodynamics:  local exchange
%  potential for nuclear structure calculations
\bibitem{SCHMID95}R.~N.\ Schmid, E.~Engel, and R.~M.\ Dreizler,
  Phys.\ Rev.\ C {\bf 52} (1995) 164.

% Local density for isovector-meson exchange
\bibitem{SCHMID95a}R.~N.\ Schmid, E.~Engel, and R.~M.\ Dreizler,
  Phys.\ Rev.\ C {\bf 52} (1995) 2804.

% Accurate density functional with correct formal properties:
%  A step beyond the generalized gradient approximation
\bibitem{PERDEW99}J.~P.\ Perdew, S.~Kurth, A.~Zupan, and P.~Blaha,
  Phys.\ Rev.\ Lett.\ {\bf 82} (1999) 2544.
  
%
\bibitem{PERDEW96}J.~P.\ Perdew, K.~Burke, and M.~Ernzerhof,
  Phys.\ Rev.\ Lett.\ {\bf 77} (1996) 3865; {\bf 78} (1997) 1396(E).

% Quantum Hadrodynamics: Evolution and Revolution
\bibitem{FURNSTAHL00}R.~J.\ Furnstahl and B.~D.\ Serot, 
    Comm.\ Nucl.\ Part.\ Phys.\ {\bf 2} (2000) A23.

% Parameter Counting in Relativistic Mean-Field Models
\bibitem{FURNSTAHL99b}R.~J.\ Furnstahl and B.~D.\ Serot,
     Nucl.\ Phys.\ {\bf A671} (2000) 447.

% Relativistic Point Coupling Models as Effective Theories of Nuclei
\bibitem{RUSNAK97}J.~J.\ Rusnak and R.~J.\ Furnstahl, 
	Nucl.\ Phys.\ {\bf A627} (1997) 495.

% The Skyrme Energy Functional and Naturalness
\bibitem{FURNSTAHL97c}R.~J.\ Furnstahl and J.~C.~Hackworth, 
     Phys.\ Rev.\ C {\bf 56} (1997) 2875.

% A chiral effective lagrangian for nuclei
\bibitem{FURNSTAHL97}R. J. Furnstahl, B. D. Serot, and H.-B. Tang,
       Nucl.\ Phys.\ {\bf A615} (1997) 441; (E) {\bf A640} (1998) 505.

% Systematics of light deformed nuclei in relativistic mean-field models
\bibitem{FURNSTAHL87}R.~J.\ Furnstahl, C.~E.\ Price, and G.~E.\
  Walker, Phys.\ Rev.\ C {\bf 36} (1987) 2590.

% The Nuclear Spin-Orbit Force in Chiral Effective Field Theories
\bibitem{FURNSTAHL98}R.~J.\ Furnstahl, J.~J.\ Rusnak, and 
    B.~D.\ Serot, Nucl.\ Phys.\ {\bf A632} (1998) 607.

% Neutron radii in nuclei and the neutron equation of state
\bibitem{BROWN00}B.~A.\ Brown, Phys.\ Rev.\ Lett.\ {\bf 85} (2000) 5296.

\bibitem{BROWN01}S.~Typel and B.~A.\ Brown, Phys.\ Rev. C {\bf 64} (2001)
  027302.

  

%  Can the equation of state of asymmetric nuclear matter be studied
%   using unstable nuclei? [RIKEN-AF-NP-267]
\bibitem{OYAMATSU}K.~Oyamatsu, I.~Tanihata, Y.~Sugahara,
K.~Sumiyoshi, and H.~Toki, Nucl.\ Phys.\ {\bf A634} (1998) 3.

% Neutron and proton densities in nuclei and the semi-empirical mass
%     formula
\bibitem{BODMER60}A.~R.\ Bodmer, Nucl.\ Phys.\ {\bf 17} (1960) 388.
%
\bibitem{CHEN99}B. Q. Chen, Z. Y. Ma,  F. Grummer, and  S.~Krewald,
  Phys.\ Lett.\ B {\bf 455} (1999) 13.
  
% Neutron density distributions for atomic parity nonconservation experiments
\bibitem{VRETENAR00}D.~Vretenar, G.~A.\ Lalazissis, and P.~Ring,
  Phys.\ Rev.\ C {\bf 62} (2000) 045502.  
   
% Masses and radii of spherical nuclei calculated in various microscopic
%  approaches
\bibitem{PATYK99}Z.~Patyk et al., Phys.\ Rev. C {\bf 59} (1999) 704.

  
% Neutron skins and halos in the mean-field theory
\bibitem{MIZUTORI00}S.~Mizutori, J.~Dobaczewski, G.~A.\ Lalazissis,
  W.~Nazarewicz, and P.-G.\ Reinhard, Phys.\ Rev.\ C {\bf 61} (2000) 044326.

% Structure of nuclei at extreme values of the isospin
\bibitem{DOBACZEWSKI99}J.~Dobaczewski, Acta Phys.\ Polon.\ B {\bf 30} (1999)
 1647 [arXiv:nucl-th/9901036].


\bibitem{KREWALD}S.~Krewald, V.~Klemt, J.~Speth, and A.~Faessler,
  Nucl.\ Phys.\ {\bf A281} (1977) 166.

        
% Ground state properties of the \beta stable nuclei in various mean
%  field theories
\bibitem{POMORSKI97}K.~Pomorski et al., Nucl.\ Phys.\ {\bf A624}
  (1997) 349.
%

% Skyrme-force parametrization: Least-squares fit to nuclear ground-state
%  properties [Skyrme 1-6, a, b, T, T3, M, M*] [E,Esig,Z,Zsig,Rsig,Gsig]
\bibitem{FRIEDRICH86}J.~Friedrich and P.-G.\ Reinhard, Phys.\ Rev.\
  C {\bf 33} (1986) 335.

% A Skyrme parametrization from subnuclear to neutron star densities
%  [SIII, SGII, SkM*, RATP, SkP, T6, Sly230a, Sly230b]
\bibitem{CHABANAT97}E.~Chabanat, P.~Bonche, P.~Haensel, J.~Meyer,
  and R.~Schaeffer, Nucl.\ Phys.\ {\bf A627} (1997) 710.


% general QHD review
\bibitem{SEROT86}B.~D.\ Serot and J.~D.\ Walecka, Adv.\ Nucl.\
    Phys.\ {\bf 16} (1986) 1.

% general QHD review
\bibitem{SEROT92} B. D. Serot, Rep.\ Prog.\ Phys.\ {\bf 55} (1992) 1855.

% QHD eft review
\bibitem{SEROT97}B.~D.\ Serot and J.~D.\ Walecka, Int.\ J.\ Mod.\
    Phys.\ E {\bf 6} (1997) 515, and references therein.

% Optimal parametrization for the realtivistic mean-field model of the nucleus
\bibitem{RUFA88}M. Rufa, P.-G.\ Reinhard, J. A. Maruhn, W. Greiner,
     and M.~R.\ Strayer, Phys.\ Rev.\  C {\bf 38} (1988) 390.

% New parametrization for the lagrangian density of relativistic mean-field
%  theory
\bibitem{LALAZISSIS97}G.~A.\ Lalazissis, J.~K\"onig, and P.~Ring,
  Phys.\ Rev.\ C {\bf 55} (1997) 540. 

%
\bibitem{NIKOLAUS97} B.\ A.\ Nikolaus, T.\ Hoch, D.\ G. Madland,
    Phys.\ Rev.\ C  {\bf 56} (1997) 177.

%
\bibitem{NEGELE70}J.~W.\ Negele, Phys.\ Rev.\ C {\bf 1}
       (1970) 1260.
%
\bibitem{NEGELE72}J.~W.\ Negele and D.~Vautherin,
        Phys.\ Rev.\ C {\bf 5} (1972) 1472.

% Large Lorentz Scalar and Vector Potentials in Nuclei
\bibitem{FURNSTAHL00b}R.~J.\ Furnstahl and B.~D.\ Serot,
     Nucl.\ Phys.\ {\bf A673} (2000) 298.

%``Relativistic Nuclear Hamiltonians,''
\bibitem{FOREST95}
J.~L.~Forest, V.~R.~Pandharipande and J.~L.~Friar,
Phys.\ Rev.\ C {\bf 52}, 568 (1995).

% Rusnak thesis
\bibitem{RUSNAKTHESIS}J.~J.\ Rusnak, Ph.D.\ thesis, 1997.


\bibitem{LIU01}B.~Liu, V.~Greco, V.~Baran, M.~Colonna, and M.~DiToro,
  arXiv:nucl-th/0112034.
    
\bibitem{BURVENICH01}
T.~Burvenich, D.~G.~Madland, J.~A.~Maruhn and P.~G.~Reinhard,
%``Nuclear ground state observables and QCD scaling in a refined  
%  relativistic point coupling model,''
arXiv:nucl-th/0111012.

\bibitem{LEPAGE89} G. P. Lepage,
  ``What is Renormalization?'', 
        in {\it From Actions to Answers} (TASI-89), edited by
        T. DeGrand and D. Toussaint (World Scientific, Singapore, 1989);
         ``How to Renormalize the Schr\"odinger Equation'',
         {\tt [nucl-th/9706029]}.
%

\bibitem{HAMMER00}
H.~W.~Hammer and R.~J.~Furnstahl,
%``Effective Field Theory for Dilute Fermi Systems,''
Nucl.\ Phys.\ A {\bf 678} (2000) 277
[arXiv:nucl-th/0004043].

\bibitem{RENTMEESTER99}
M.~C.~Rentmeester, R.~G.~Timmermans, J.~L.~Friar and J.~J.~de Swart,
%``Chiral two-pion exchange and proton proton partial-wave analysis,''
Phys.\ Rev.\ Lett.\  {\bf 82} (1999) 4992.


  
\bibitem{JACKSON92}A.~D.\ Jackson, in {\it Recent Progress in
       Many-Body Theories,\/} vol.~3, ed.~T.~L.\ Ainsworth,
	    C.~E.\ Campbell, B.~E.\ Clements, and E.~Krotscheck
		(Plenum, 1992).

\bibitem{JACKSON94}A.~D.\ Jackson and T.~Wettig, Phys.\ Rep.\ {\bf 237}
  (1994) 325.

% chiral quarks and the nonrelativistic quark model
\bibitem{GEORGI84b}H. Georgi and A. Manohar,
       Nucl.\ Phys.\ {\bf B234} (1984) 189.

% Generalized dimensional analysis
\bibitem{GEORGI93b}H.\ Georgi, Phys. Lett. {\bf B298} (1993) 187.

%
% VMD
%
%
% Friar scales
\bibitem{FRIAR96b}J.~L.\ Friar, Few Body Syst.\ {\bf 22} (1997) 161.
%
\bibitem{FRIAR96}J. L. Friar, D. G. Madland, and B. W. Lynn,
        Phys.\ Rev.\  C {\bf 53} (1996) 3085.
  
\bibitem{HOROWITZ81}C.~J.~Horowitz and B.~D.~Serot,
%``Selfconsistent Hartree Description Of Finite Nuclei In A 
% Relativistic Quantum Field Theory,''
Nucl.\ Phys.\ A {\bf 368}, 503 (1981).

\bibitem{MUELLER96}H.~Mueller and B.~D.~Serot,
%``Relativistic Mean-Field Theory and the High-Density Nuclear 
%  Equation of State,''
Nucl.\ Phys.\ A {\bf 606}, 508 (1996)

%


\bibitem{FURNSTAHL01a}R.~J.~Furnstahl and H.~W.~Hammer,
%``Are Occupation Numbers Observable?,''
arXiv:nucl-th/0108069.

%
\bibitem{LEE98}C.-H.\ Lee, T.~T.~S.\ Kuo, G.~Q.\ Li, and G.~E.\ Brown, 
   Phys.\ Rev.\ C {\bf 57} (1998) 3488.

\bibitem{HOROWITZPC}C.~J.\ Horowitz, private communication.
  
% Modern Nucleon-nucleon potentials and symmetry energy in infinite matter
\bibitem{ENGVIK97}L.~Engvik, M.~Hjorth-Jensen, R.~Machleidt, H.~Muther,
 and A.~Polls, Nucl.\ Phys.\ {\bf A627} (1997) 85.

% Neutron star profiles in the relativistic bruckner-hartree-fock theory
\bibitem{SUMIYOSHI95}K.~Sumiyoshi, K.~Oyamatsu, and H.~Toki,
  Nucl.\ Phys.\ {\bf A595} (1995) 327.
 


\bibitem{HU99}Y.~Hu, Ph.D.\ thesis, Indiana University, 2000.


\bibitem{PUGLIA}R.~J.\ Furnstahl, H.-W.\ Hammer, and S.~Puglia,
   in preparation.

\bibitem{KARATAGLIDIS01}
S.~Karataglidis, K.~Amos, B.~A.~Brown and P.~K.~Deb,
%``Discerning the neutron density distribution of 208Pb from nucleon elastic scattering,''
arXiv:nucl-th/0111020.


\bibitem{BUNNY}B.~C.\ Clark and S.~Hama, in preparation.

\bibitem{TRZCINSKA01}A.~Trzcinska, J.~Jastrzebski, P.~Lubinski,
   F.~J.\ Hartmann, R.~Schmidt, T.\ von Egidy, and B.~Klos,
   Phys.\ Rev.\ Lett.\ {\bf 87} (2001) 082501.

% Nuclear surface properties and spin orbit potential in a modified derivative
%  coupling model
%\bibitem{HUA99}G. Hua, T.~v.\ Chossy, and W.~Stocker, Phys.\ Rev.\ 
%C {\bf 61} (2000) 014307.

%\bibitem{DREIZLER00}C.~Speicher, R.~M.\ Dreizler, and E.~Engel, 
%   Ann.\ Phys.\ (N.Y.) {\bf 213} (1992) 312.
%
% OBEP
%

% Relativistic and nonrelativistic impulse approximation descriptions of
%  300-1000 MeV proton + nucleus elastic scattering
%\bibitem{RAY85}L.~Ray and G.~W.\ Hoffmann, Phys.\ Rev.\ C {\bf 31} (1985) 538.



%\bibitem{POLLOCK99}Pollock and Welliver, Phys.\ Lett.\ {\bf B464}
%  (1999) 177.



\end{thebibliography}
\end{document}